\documentclass[fleqn,usenatbib]{mnras}

\usepackage[T1]{fontenc}

\usepackage{graphicx}	
\usepackage{amsmath}	
\usepackage[utf8]{inputenc}
\usepackage{xcolor}
\usepackage{newtxtext,newtxmath}

\title[Effects of dust particles on Alfvén waves]{Effects of dust particles charged by inelastic collisions and by photoionization on Alfvén waves in a stellar wind}

\author[L. B. De Toni, R. Gaelzer]{
L. B. De Toni,$^{1}$\thanks{E-mail: luan.toni@ufrgs.br}
R. Gaelzer,$^{1}$\thanks{E-mail: rudi.gaelzer@ufrgs.br}
\\
$^{1}$Instituto de Física, Universidade Federal do Rio Grande do Sul, CP 15051, 91501-970, Porto Alegre, RS, Brazil
}

\date{This is a pre-copyedited, author-produced PDF of an article accepted for publication in Monthly Notices of the Royal Astronomical Society following peer review.}

\pubyear{2021}

\begin{document}
\label{firstpage}
\pagerange{\pageref{firstpage}--\pageref{lastpage}}
\maketitle

\begin{abstract}
Using a kinetic description of a homogeneous magnetized dusty plasma with Maxwellian distribution of electrons and protons and dust particles charged by inelastic collisions and by photoionization, we analyse the dispersion relation considering the case where waves and radiation propagate exactly parallel to the ambient magnetic field. The investigation emphasizes the changes that the photoionization process brings to the propagation and damping of the waves in a stellar wind environment, since Alfvén waves are believed to play a significant role in the heating and acceleration processes that take place in the wind. The results show that, in the presence of dust with negative equilibrium electrical charge, the Alfvén mode decouples into the whistler and ion cyclotron modes for all values of wavenumber, but when dust particles acquire neutral or positive values of electrical charge, these modes may couple for certain values of wavenumber. It is also seen that the whistler and ion cyclotron modes present null group velocity in a interval of small wavenumber, and that the maximum value of wavenumber for which the waves are non-propagating is reduced in the presence of the photoionization process. For very small values of wavenumber, the damping rates of the modes could change significantly from very small to very high values if the sign of the dust electrical charge is changed. 
\end{abstract}




\section{Introduction}

Dust particles are commonly present in various space environments, such as stellar winds, planetary rings, planetary atmospheres, comet tails and interstellar medium, which are also composed by a fully ionized plasma composed by electrons and ions. The combination of a traditional plasma embedded with charged dust particles can be defined as \emph{dust in a plasma} (where the mean distance between dust grains is greater than the Debye length and they can be treated as isolated particles) or as a \emph{dusty plasma} (where the charged dust particles participate in the collective behavior).

Once a dust particle is embedded in a plasma it acquires an electrical charge from a variety of charging mechanisms such as absorption of charged particles by inelastic collisions, photoionization, secondary electron emission, field emission, among others. For a more detailed account of the main charging mechanisms of dust grains see e.g. \protect\citet{shukla_book}. 

This variable charge modifies the dielectric properties of the dusty plasma. In particular, it may change the dispersion relations of normal wave modes causing the appearance of additional damping mechanisms and may also be responsible for the emergence of new wave modes, such as the dust acoustic (DA) waves, dust ion acoustic (DIA) waves and electrostatic dust ion cyclotron (EDIC) waves \citep{Merlino_1998_laboratory,DANGELO_1990,Rao_1990}.

Of particular importance is the study of damping and propagation of Alfvén waves in stellar winds. These winds are responsible for the mass loss of massive stars and their acceleration may be caused by several mechanisms, being Alfvén waves proposed as one of them. The idea that Alfvén waves are present in the winds of stars in many regions of the HR diagram is supported by the fact that these waves are observed in the solar wind \citep{Smith_1995Ulysses,Tomczyk_2007Alfven} and they are generated by oscillations in the magnetic field at the base of the wind. As they propagate, these waves dissipate energy and transfer momentum to the plasma, which can heat and accelerate the wind.

Alfvénic fluctuations are ubiquitous in the solar wind and are related to important processes that take place there and are recorded by \emph{in situ} observations.  
For instance, Alfvén waves may be related to the observed temperatures of the different particle species present in the solar wind, particularly to the proton velocity distribution, which displays a "double-humped" feature, with peaks separated by the local Alfvén speed \citep{Marsch06}\@. Low-frequency Alfvénic turbulence may also be chiefly responsible (via several nonlinear processes) for the (larger) observed profile of magnetic field turbulence in both the fast and slow solar winds.  Moreover, there is growing evidence that the decades-old enigma of the mechanism responsible for the heating of the solar corona and acceleration of the solar wind is related to nonlinear kinetic processes that take place between the local plasma and upward-propagating (from the solar chromosphere) Alfvén waves \citep{DePontieu+07/12, Cranmer+15/04}\@.  Rather than giving a long list of publications, the Reader is referred to the recent compilation by \citet{Raouafi+21/04}.

\citet{jatenco1989effect} proposed a model for mass loss in late-type stars where they used a flux of Alfvén waves as an acceleration mechanism of the wind, considering wave damping due to nonlinear effects, surface Alfvén wave absorption and turbulent effects. Later, \citet{Falceta_Goncalves_2002} included in the model of \citet{jatenco1989effect} the effects of radiation pressure on grains and a new strong damping mechanism of Alfvén waves due to the presence of dust, resulting in a more realistic (and consistent with observations) acceleration mechanism of winds of cool supergiant stars.

Several works have been published studying the effects that dust particles cause in the propagation and damping of waves in a stellar environment using a kinetic formulation and considering only the absorption of plasma particles as the charging mechanism \citep{dejuli_schneider_1998,Juli_2005,Ziebell_2005,schneider_2006_electrostatic,deJuli_2007effect,deJuli_2007mode,ziebell_2008,Gaelzer_2008,Gaelzer_2010}. Some of the results show that the presence of dust particles with variable charge leads to several changes on the dynamic properties of the waves; for example, the appearance of an additional damping mechanism of Alfvén waves beyond the conventional Landau damping, and the occurrence of mode coupling between distinct branches of the dispersion relation.

More recently, \citet{galvao_ziebell2012} developed the formalism for a kinetic theory for magnetized dusty plasma, including the photoionization as charging mechanism of the dust particles together with absorption of plasma particles. In this work, we make use of the formulation of \citet{galvao_ziebell2012} to analyse the modification that this new charging process brings to the propagation and damping of Alfvén waves that propagate parallel to the ambient magnetic field.

As we will see, the addition of the photoionization process can change the real part of the frequencies of the whistler and ion cyclotron modes for small values of wavenumber, reducing the region of null group velocity, a feature observed in the presence of dust in the plasma. This new charging mechanism could also change significantly the imaginary parts, i.e., the damping rates of these modes for very small values of wavenumber if the equilibrium electrical charge of the dust particles acquires a null or positive value, which becomes possible when photoionization is considered.

The plan of this paper is as follows. In Section~\ref{sec:The-Model} we describe the basic features of the dusty plasma model employed and the charging processes considered. Section~\ref{sec:Dispersion-Relation} presents the dispersion relation for Alfvén waves with Maxwellian distributions for plasma particles. In Section~\ref{sec:Numerical-Results} we present some numerical results of the dispersion relation considering parameters typically found in stellar winds. The conclusions are presented in Section~\ref{sec:Conclusions}. In the Appendix, we evaluate the component of the dielectric tensor that arises from the addition of the photoionization process, within the approximations considered in this work.

\section{The Model}
\label{sec:The-Model}

The model we use in this work considers a homogeneous plasma composed by electrons, protons and spherical dust particles in the presence of an ambient magnetic field $\mathbfit{B}_{0}=B_{0}\mathbfit{e}_{z}$, in an environment with incidence of anisotropic radiation. In this model, the dust grains have constant radius $a$ and variable charge $q_\mathrm{d}$, which originates from inelastic collisions between the dust particles and particles of species $\beta$ (electrons and ions) with charge $q_{\beta}$ and mass $m_{\beta}$, and from the emission of electrons by photoionization.

The dusty plasma is considered collisionless except for the inelastic collisions that lead to the charging of the dust grains. Therefore, the model considers a transport equation for the plasma particles that includes a collision term due to the absorption of plasma particles by the dust grains and a source term related to emitted photoelectrons. To acknowledge the variation of the dust electrical charge in a kinetic description of the dusty plasma system, the dust charge is treated as a dynamical variable and a kinetic equation for the dust particles is considered. A detailed account of this formalism is given by \citet{vladimirov_1994}, \citet{dejuli_schneider_1998} and \citet{galvao_ziebell2012}.

Since all dust particles are assumed to have the same mass, which is much larger than the masses of either protons or electrons, they are assumed to be immobile. Consequently, this model will be restricted to waves with frequency much higher than the characteristic dust frequencies, thereby excluding the modes that can arise from the dust dynamics. That is, we consider the regime in which $\omega\gg\max\left(\omega_\mathrm{d},|\Omega_\mathrm{d}|\right)$, where $\omega_\mathrm{d}$ and $\Omega_\mathrm{d}$ are, respectively, the plasma and cyclotron frequencies of the dust particles.

\subsection{Dust charging processes}
\label{subsec:Dust-Charging-Processes}

Given that dust particles are considered immobile, we will work in a range of frequencies well above the characteristic frequencies associated with the motion of the dust particles so that we may assume that the electric charge deposited on the grain's surface have reached an equilibrium value. To evaluate this equilibrium dust charge we use the condition of zero surface current,
\begin{equation}
    I_{0}(q_\mathrm{d0}=Z_\mathrm{d}e)=0,
    \label{eq:I_0=00003D0}
\end{equation}
where $q_\mathrm{d0}$ is the equilibrium dust charge, $Z_\mathrm{d}$ is the dust charge number, $e$ is the elementary charge and $I_{0}$ is the charging current over the surface of a dust grain, which is given by the zeroth-order solution of the linearized Vlasov-Maxwell set of equations and has the following expression \citep{galvao_ziebell2012}:
\begin{equation}
    I_{0}(q_\mathrm{d})=\sum_{\beta}I_{\beta0}(q_\mathrm{d})+I_\mathrm{p}(q_\mathrm{d}),
\end{equation}
where $I_{\beta0}$ is the current due to absorption of plasma particles of species $\beta$ in the equilibrium, and $I_\mathrm{p}$ is the photoemission current.

Because dust particles acquire an equilibrium charge, the equilibrium number densities of the different species of plasma particles are no longer the same, as expected in a two-species plasma. Instead, the equilibrium number densities are obtained from the quasi-neutrality condition
\begin{equation}
    \sum_{\beta}n_{\beta0}q_{\beta}+q_\mathrm{d0}n_{d0}=0,
    \label{eq:quasineutrality}
\end{equation}
where $n_{\beta0}$ and $n_{d0}$ are the equilibrium density of the plasma particles of species $\beta$ and the dust particles, respectively.

The absorption current is caused by inelastic collisions of plasma particles with dust particles, and has a cross section described by the orbital motion limited (OML) theory \citep[see e.g.][]{Allen_1992,Tsytovich_1997}, which neglects the presence of the ambient magnetic field in the charging process. In principle, the magnetic field should influence the absorption current since it changes the motions of plasma particles. However, a numerical calculation performed by \citet{Chang_1993} shows that the charging of the dust particles by inelastic collisions is not significantly influenced by the presence of an external magnetic field when $a\ll\rho_\mathrm{G}$, where $\rho_\mathrm{G}=(\pi/2)^{1/2}r_{Le}$ and $r_{Le}$ is the electron Larmor radius. 

\citet{salimullah2003dust} used a modified OML theory to show that, in a strongly magnetized dusty plasma, 
the dust charging process and the damping of low-frequency electrostatic waves are modified
when the plasma species' Larmor radius is comparable to the dust particle radius. This effect happens because the orbits of plasma particles are confined to one dimension along the magnetic field lines, making the dusty plasma anisotropic. Hence, charging currents to a spherical grain are not the same in distinct directions.

More recently, \citet{Kodanova+2019} used a particle-in-cell simulation to calculate the dust particle charge at different values of the magnetic field. Their work shows that an external magnetic field could reduce significantly the absolute value of the dust particle charge when the magnetic field is sufficiently strong, resulting in an electron Larmor radius smaller than the dust particle radius. On the other hand, for values of magnetic field where the electron Larmor radius is much larger than the dust radius, the value of the dust electrical charge is quite similar to that where no magnetic field is considered, corroborating the work of \citet{Chang_1993}. 

For the values of parameters used in this work, the relation $a\ll\rho_\mathrm{G}$ is always satisfied, i.e., the electron Larmor radius is always much greater than the dust particle radius, making it possible to use the OML theory in our formulation.

Therefore, the explicit expression of the absorption current is given by \citep{dejuli_schneider_1998}
\begin{equation}
    I_{\beta0}(q_\mathrm{d})=\pi a^{2}q_{\beta}\int \mathrm{d}^{3}p\left(1-\frac{C_\beta}{p^{2}}\right)H\left(1-\frac{C_\beta}{p^{2}}\right)\frac{p}{m_{\beta}}f_{\beta0},
    \label{eq:I_beta0_root}
\end{equation}
where
\begin{equation}
    C_\beta \equiv \frac{2 q_d q_\beta m_\beta }{a},    
\end{equation}
and $p$ is the momentum of the plasma particles, $f_{\beta0}$ is the distribution function of species $\beta$ in equilibrium and $H\left(x\right)$ is the Heaviside function. Assuming Maxwellian distribution for the plasma particles, equation~\eqref{eq:I_beta0_root} can be evaluated for electrons ($\beta=e$), resulting
\begin{equation}
    I_{e0}(q_\mathrm{d})=-2\sqrt{2\pi}a^{2}en_{e0}v_{Te}
    \begin{cases}
        \exp\left(\frac{q_\mathrm{d}e}{ak_\mathrm{B}T_{e}}\right), &q_\mathrm{d}<0\\
        \left(1+\frac{q_\mathrm{d}e}{ak_\mathrm{B}T_{e}}\right), &q_\mathrm{d}\geq0
    \end{cases},
\end{equation}
and for protons ($\beta=i$), resulting
\begin{equation}
    I_{i0}(q_\mathrm{d})=2\sqrt{2\pi}a^{2}en_{i0}v_{Ti}
    \begin{cases}
        \left(1-\frac{q_\mathrm{d}e}{ak_\mathrm{B}T_{i}}\right), &q_\mathrm{d}\leq0\\
        \exp\left(-\frac{q_\mathrm{d}e}{ak_\mathrm{B}T_{i}}\right), &q_\mathrm{d}>0
    \end{cases},
\end{equation}
where $k_\mathrm{B}$ is the Boltzmann constant, and $v_{T\beta}=(k_\mathrm{B}T_{\beta}/m_{\beta})^{1/2}$ and $T_{\beta}$ are, respectively, the thermal velocity and temperature of the plasma particles of species $\beta$.

Since the electron thermal speed is much larger than the ion thermal speed, the absorption of ions by the dust particles will be negligible when compared with the absorption of electrons. So, we may consider a constant ion density in order to evaluate the equilibrium dust charge together with the electron density with equations~\eqref{eq:I_0=00003D0} and \eqref{eq:quasineutrality}.

The model that describes the photoelectric current assumes that electrons at the surface of the dust grains have a certain probability to absorb the incoming radiation and, when absorbed, these electrons can be emitted if the energy of the radiation is greater than the work function of the material of the grain. For the case of a positively charged dust particle, the energy of the emitted electron must overcome the electrostatic attraction by the grain, otherwise it will be reabsorbed by the dust particle. This model also assumes that the number of electrons emitted by unit area by unit time is proportional to the intensity of radiation, and the distribution of momenta of the electrons in the material obeys the Fermi-Dirac statistics.

For the case in which radiation is unidirectional, propagating in parallel with the ambient magnetic field, the photoelectric current for a spherical dust grain with charge $q_\mathrm{d}$ uniformly distributed over its surface can be written as \citep{galvao_ziebell2012}
\begin{equation}
    I_\mathrm{p}=\frac{2}{h^{3}}\int \mathrm{d}^{3}p\sigma_\mathrm{p}(p,q_\mathrm{d})\frac{p_{z}}{m_{e}}\left[1+\exp\left(\frac{p^{2}}{2m_{e}k_\mathrm{B}T_\mathrm{d}}-\xi\right)\right]^{-1},
    \label{eq:I_p_integral}
\end{equation}
where $h$ is the Planck constant, $T_\mathrm{d}$ is the dust temperature,
\begin{equation}
    \xi=\frac{1}{k_\mathrm{B}T_\mathrm{d}}(h\nu-\phi)
\end{equation}
with $\nu$ being the incident radiation frequency, and $\phi$ the
work function of the material. We also have
\begin{equation}
    \sigma_\mathrm{p}(p,q_\mathrm{d})=e\pi a^{2}\beta(\nu)\Lambda(\nu)S_{a}H(p_{z})H\left(1+\frac{C_e}{p^{2}}H(q_\mathrm{d})\right),
\end{equation}
which is the photoemission cross section, where $C_e = -2m_e e q_d/a$, $\beta(\nu)$ is the probability of an electron which arrives to the surface coming from the inside to absorb a photon of frequency $\nu$ at the surface, $\Lambda(\nu)$ is the number of photons with frequency $\nu$ incident per unit of area per unit of time, and $S_{a}=S_{e}-S_{s}$ where $S_{e}$ and $S_{s}$ are, respectively, the extinction and scattering coefficients, accordingly with Mie theory \citep[see e.g.][]{Sodha_2011}. We consider $S_{a}=1$ which is a fair approximation when $2\pi a/\lambda\geq10$, where $\lambda$ is the wavelength of the radiation. This condition is always satisfied within the parameters used in this work.

Evaluating the integral in equation~\eqref{eq:I_p_integral}, we have:
\begin{equation}
    I_\mathrm{p}=e a^{2}\frac{4\pi^2 m_{e}(k_\mathrm{B}T_\mathrm{d})^{2}}{h^{3}}\beta(\nu)\Lambda(\nu)S_{a}
    \begin{cases}
        \Phi(\xi), &q_\mathrm{d}\leq0\\
        \Psi(\xi,q_\mathrm{d}), &q_\mathrm{d}>0
    \end{cases},
\end{equation}
where
\begin{equation}
    \Phi(\xi)=\int_{0}^{\exp\xi} \frac{\ln(1+\Omega)}{\Omega} \mathrm{d}\Omega=-\text{Li}_{2}(-\exp\xi),
\end{equation}
\begin{equation}
\begin{aligned}
    \Psi(\xi,q_\mathrm{d})=&\frac{eq_\mathrm{d}}{ak_\mathrm{B}T_\mathrm{d}}\ln\left[1+\exp\left(\xi-\frac{eq_\mathrm{d}}{ak_\mathrm{B}T_\mathrm{d}}\right)\right]\\
    &+\Phi\left(\xi-\frac{eq_\mathrm{d}}{ak_\mathrm{B}T_\mathrm{d}}\right),
\end{aligned}
\end{equation}
and $\text{Li}_{2}$ is the polylogarithm function of order $2$.

Now we define the photoelectric efficiency of the dust material, or the fraction of absorbed photons that leads to the emission of a photoelectron,
as \citep[see][]{Sodha_2009}
\begin{equation}
    \chi(\nu)=\frac{4\pi m_{e}(k_\mathrm{B}T_\mathrm{d})^{2}}{h^{3}}\beta(\nu)\Phi(\xi).
\end{equation}
The spectral dependence of $\chi(\nu)$ was formulated by \citet{Spitzer_1948} from a study of a number of photoemitters and it can be expressed as
\begin{equation}
    \chi(\nu)=
    \begin{cases}
        0, &\nu<\nu_{0}\\
        \frac{729}{16}\left(\frac{\nu_{0}}{\nu}\right)^{4}\left(1-\frac{\nu_{0}}{\nu}\right)\chi_\mathrm{m}, &\nu>\nu_{0}
    \end{cases},
\end{equation}
where $\nu_{0}=\phi/h$ is the threshold frequency and $\chi_\mathrm{m}$    is the maximum value of the photoelectric efficiency.

Expressing the photoelectric current with this new parameter we have
\begin{equation}
\begin{alignedat}{2} 
    & I_\mathrm{p}=e\pi a^{2}S_{a}\chi(\nu)\Lambda(\nu), &q_\mathrm{d}\leq0,\\
    & I_\mathrm{p}=e\pi a^{2}S_{a}\frac{\Psi(\xi,q_\mathrm{d})}{\Phi(\xi)}\chi(\nu)\Lambda(\nu), \quad&q_\mathrm{d}>0.
\end{alignedat}
\label{eq:I_p_mono}
\end{equation}

For a continuous spectrum, equation~\eqref{eq:I_p_mono} can be generalised \citep[see][]{Sodha_1963} to 
\begin{equation}
\begin{alignedat}{2} 
    & I_\mathrm{p}=e\pi a^{2}\int_{\nu_{0}}^{\nu_\mathrm{m}}S_{a}\chi(\nu)\Lambda(\nu) \mathrm{d}\nu, &q_\mathrm{d}\leq0\,,\\
    & I_\mathrm{p}=e\pi a^{2}\int_{\nu_{0}}^{\nu_\mathrm{m}}S_{a}\frac{\Psi(\xi,q_\mathrm{d})}{\Phi(\xi)}\chi(\nu)\Lambda(\nu) \mathrm{d}\nu, \quad&q_\mathrm{d}>0\,,
\end{alignedat}
\label{eq:I_p_continuous}
\end{equation}
where $\nu_\mathrm{m}$ is the upper limit of the spectrum defined by either $\chi(\nu_\mathrm{m})\approx0$ or $\Lambda(\nu_\mathrm{m})\approx0$, and $\Lambda(\nu)\mathrm{d}\nu$ is the number of photons with frequency between $\nu$ and $\nu+\mathrm{d}\nu$ incident per unit time per unit area.

For a stellar environment, we may consider that the star radiate as a black body, so that \citep[see e.g.][]{Srivastava_2016}
\begin{equation}
    \Lambda(\nu)\mathrm{d}\nu=\frac{4\pi\nu^{2}}{c^{2}}\left[\exp\left(\frac{h\nu}{k_\mathrm{B}T_\mathrm{s}}\right)-1\right]^{-1}\left(\frac{r_\mathrm{s}}{r_\mathrm{d}}\right)^{2}\mathrm{d}\nu,
    \label{eq:blackbody}
\end{equation}
where $c$ is the speed of light in vacuum, $T_\mathrm{s}$ is the surface temperature of the star, $r_\mathrm{s}$ is the radius of its radiating surface, and $r_\mathrm{d}$ is the mean distance of the dust grains from the star.

\section{Dispersion Relation for Alfvén Waves}
\label{sec:Dispersion-Relation}

The components of the dielectric tensor for a homogeneous magnetized dusty plasma, fully ionized, with immobile dust particles and charge variable in time, can be written in the following way:
\begin{equation}
    \epsilon_{ij}=\epsilon_{ij}^\mathrm{C}+\epsilon_{ij}^\mathrm{A}+\epsilon_{ij}^\mathrm{P},
\end{equation} for $\left\{ i,j\right\} =\left\{ 1,2,3\right\} $, where explicit expressions for $\epsilon_{ij}^\mathrm{C}$, $\epsilon_{ij}^\mathrm{A}$ and $\epsilon_{ij}^\mathrm{P}$ are given e.g. by \citet{dejuli_schneider_1998} and \citet{galvao_ziebell2012}.

The terms $\epsilon_{ij}^\mathrm{C}$ refer to the dielectric tensor components of a conventional (dustless) plasma of electrons and ions, since they have the same formal structure, except for the $i3$ components and the addition of a purely imaginary term in the resonant denominator. The terms $\epsilon_{ij}^\mathrm{A}$ and $\epsilon_{ij}^\mathrm{P}$ are entirely new and only exist in the presence of dust particles. The former is associated with the current due to absorption of plasma particles, while the latter is associated with the current due to photoelectric emission.

It has been shown that the additional $i3$ components appearing in $\epsilon_{ij}^\mathrm{C}$ due to the absorption current vanishes for Maxwellian distribution and, for propagation parallel to the magnetic field, the term $\epsilon_{ij}^\mathrm{A}$ only occurs for $i=j=3$, regardless of the form of the distribution function \citep[see][]{Juli_2005}. Likewise, is easy to show that the term $\epsilon_{ij}^\mathrm{P}$ also only occurs for $i=j=3$ when considering parallel propagation of the wave together with incidence of the radiation also parallel to the magnetic field (see equation~\eqref{eq:tensor_photo}).

Therefore, considering propagation parallel to the ambient magnetic field, with electrons and protons described by a Maxwellian distribution, a dust population with constant radius $a$, and radiation also propagating in parallel to the magnetic field, the dielectric tensor assumes the form

\renewcommand\arraystretch{1.5}
\begin{equation}
    \mathbf{\epsilon} =\left(\begin{array}{ccc}
    \epsilon_{11}^\mathrm{C} & \epsilon_{12}^\mathrm{C} & 0\\
    -\epsilon_{12}^\mathrm{C} & \epsilon_{11}^\mathrm{C} & 0\\
    0 & 0 & \epsilon_{33}^\mathrm{C}+\epsilon_{33}^\mathrm{A}+\epsilon_{33}^\mathrm{P}
    \end{array}\right),
\end{equation}
where
\begin{align}
    &\epsilon_{11}^\mathrm{C}=1+\frac{1}{4}\sum_{\beta}X_{\beta}\left(\hat{I}_{\beta}^{+}+\hat{I}_{\beta}^{-}\right),\\
    &\epsilon_{12}^\mathrm{C}=-\frac{i}{4}\sum_{\beta}X_{\beta}\left(\hat{I}_{\beta}^{+}-\hat{I}_{\beta}^{-}\right),
\end{align}
and where
\begin{align}
    \hat{I}_{\beta}^{s}=\frac{1}{n_{\beta0}}\int \mathrm{d}^{3}p\frac{p_{\perp}\partial f_{\beta0}/\partial p_{\perp}}{1-\frac{k_{\parallel}p_{\parallel}}{m_{\beta}\omega}+s\frac{\Omega_{\beta}}{\omega}+\mathrm{i}\frac{\nu_{\beta d}^{0}(p)}{\omega}},
\end{align}
with
\begin{align}
   & \nu_{\beta d}^{0}(p)=\frac{\pi a^{2}n_{d0}}{m_{\beta}}\frac{\left(p^{2}-\frac{2m_{\beta}q_{\beta}q_\mathrm{d0}}{a}\right)}{p}H\left(p^{2}-\frac{2m_{\beta}q_{\beta}q_\mathrm{d0}}{a}\right),\\
   & X_{\beta}=\frac{\omega_{p\beta}^{2}}{\omega^{2}},\quad\omega_{p\beta}^{2}=\frac{4\pi n_{\beta0}q_{\beta}^{2}}{m_{\beta}},\quad\Omega_{\beta}=\frac{q_{\beta}B_{0}}{m_{\beta}c},
\end{align}
and $s=\pm1$. The explicit form of the $\epsilon_{33}$ component will not be reproduced here since it does not contribute to the dispersion relation of Alfvén waves, in which we focus our attention in this work.

The general dispersion relation for $\mathbfit{k}=k_{\parallel}\mathbfit{e}_{z}$ follows from the equation
\begin{equation}
    \det\begin{pmatrix}\epsilon_{11}^\mathrm{C}-N_{\text{\ensuremath{\parallel}}}^{2} & \epsilon_{12}^\mathrm{C} & 0\\
    -\epsilon_{12}^\mathrm{C} & \epsilon_{11}^\mathrm{C}-N_{\text{\ensuremath{\parallel}}}^{2} & 0\\
    0 & 0 & \epsilon_{33}^\mathrm{C}+\epsilon_{33}^\mathrm{A}+\epsilon_{33}^\mathrm{P}
    \end{pmatrix}=0,\label{eq:Dispersion-equation}
\end{equation}
where $N_{\parallel}=k_{\parallel}c/\omega$ is the refractive index in the parallel direction to the ambient magnetic field. The dispersion relation for Alfvén waves is obtained by ignoring the longitudinal mode, that is, by imposing $E_{3}=0$. As a result, we obtain
\begin{equation}
    \left[N_{\parallel}^{2}\right]_{s}=1+\frac{1}{2}\sum_{\beta}X_{\beta}\hat{I}_{\beta}^{s},
\end{equation}
where $s=+\left(-\right)1$ identifies the right (left) circularly-polarized eigenmode.

To evaluate the integrals $\hat{I}_{\beta}^{s}$ we follow the same procedure used in previous works \citep{Juli_2005,Ziebell_2005,deJuli_2007mode,Gaelzer_2010}, where the momentum dependent inelastic collision frequency $\nu_{\beta d}^{0}(p)$ is replaced by its average values in momentum space, 
\begin{equation}
    \nu_{\beta}=\frac{1}{n_{\beta0}}\int \mathrm{d}^{3}p\,\nu_{\beta d}^{0}(p)f_{\beta0}.
\end{equation}
This approximation is adopted in order to arrive at a relative simple expression for the dispersion relation and has been validated by a study performed by \citet{deJuli_2007mode} that demonstrated that in the Alfvén wave frequency range the dispersion relation can be satisfyingly described by the average collision frequency approximation. For Maxwellian distributions we obtain
\begin{align}
    & \nu_{i}=2\sqrt{2\pi}a^{2}n_{d0}v_{Ti}
    \begin{cases}
        \left(1-\frac{eq_\mathrm{d0}}{ak_\mathrm{B}T_{i}}\right), &q_\mathrm{d0}\leq0\\
        \exp\left(-\frac{eq_\mathrm{d0}}{ak_\mathrm{B}T_{i}}\right), &q_\mathrm{d0}>0
    \end{cases},\\
    & \nu_{e}=2\sqrt{2\pi}a^{2}n_{d0}v_{Te}
    \begin{cases}
        \exp\left(\frac{eq_\mathrm{d0}}{ak_\mathrm{B}T_{e}}\right), &q_\mathrm{d0}<0\\
        \left(1+\frac{eq_\mathrm{d0}}{ak_\mathrm{B}T_{e}}\right), &q_\mathrm{d0}\geq0
    \end{cases}.
\end{align}

Using these expressions to evaluate the integrals $\hat{I}_{\beta}^{s}$ we obtain for the dispersion relation:
\begin{equation}
    \left[N_{\parallel}^{2}\right]_{s}=1+\sum_{\beta}X_{\beta}\zeta_{\beta}^{0}Z(\hat{\zeta}_{\beta}^{s}),\label{eq:disp_rel_dim}
\end{equation}
where 
\begin{equation}
    \zeta_{\beta}^{0}=\frac{\omega}{\sqrt{2}k_{\parallel}v_{T\beta}},\quad\hat{\zeta}_{\beta}^{s}=\frac{\omega+s\Omega_{\beta}+\mathrm{i}\nu_{\beta}}{\sqrt{2}k_{\parallel}v_{T\beta}},
\end{equation}
and $Z$ is the plasma dispersion function \citep[see][]{FriedContebook1961}, given by 
\begin{equation}
    Z(\zeta)=\frac{1}{\sqrt{\pi}}\int_{-\infty}^{+\infty}\frac{e^{-t^{2}}}{t-\zeta}\mathrm{d}t.
\end{equation}

The equation~\eqref{eq:disp_rel_dim} is formally the same dispersion relation for Alfvén waves evaluated in previous works \citep{Juli_2005,Ziebell_2005}, where only the absorption of particles is taken into account as charging mechanism of the dust particles. Therefore, the only modification that the addition of the photoelectric current brings to this dispersion relation is related to the equilibrium dust charge $q_\mathrm{d0}$, which may modify the electron density via quasi-neutrality condition (see equation~\eqref{eq:quasineutrality}) if we fix the ion and dust densities, and may also change the characteristics of the average inelastic collision frequency $\nu_{\beta}$ since it is dependent of the equilibrium dust charge.

In order to evaluate the numerical solution of the dispersion relation, we introduce the following dimensionless quantities:
\begin{equation}
\begin{alignedat}{2}
    &z=\frac{\omega}{\Omega_{i}},\quad \varepsilon=\frac{n_{d0}}{n_{i0}},\quad  u_{\beta}=\frac{v_{T\beta}}{v_\mathrm{A}},\quad \chi_{\beta}=\frac{q_\mathrm{d0}q_{\beta}}{ak_\mathrm{B}T_{\beta}},\\
    &\gamma=\frac{\lambda^{2}n_{i0}v_\mathrm{A}}{\Omega_{i}},\quad  \tilde{a}=\frac{a}{\lambda},\quad \lambda=\frac{e^{2}}{k_\mathrm{B}T_{i}},\quad  q=\frac{k_{\parallel}v_\mathrm{A}}{\Omega_{i}},\\
    &\tilde{\nu}_{\beta}=\frac{\nu_{\beta}}{\Omega_{i}},\quad \eta_{\beta}=\frac{\omega_{p\beta}}{\Omega_{i}},\quad  r_{\beta}=\frac{\Omega_{\beta}}{\Omega_{i}},
\end{alignedat}
\end{equation}
where $v_\mathrm{A}$ is the Alfvén velocity,
\begin{equation}
    v_\mathrm{A}^{2}=\frac{B_{0}^{2}}{4\pi n_{i0}m_{i}}.
\end{equation}

In terms of these quantities, the collision frequencies are given
by the following dimensionless expression
\begin{equation}
    \tilde{\nu}_{\beta}=2\sqrt{2\pi}\varepsilon\gamma\tilde{a}^{2}u_{\beta}
    \begin{cases}
        \left(1+|\chi_{\beta}|\right), &\chi_{\beta}\leq0\\
        \exp\left(-\chi_{\beta}\right), &\chi_{\beta}>0
    \end{cases},
    \label{eq:colision_freq}
\end{equation}
and the dispersion relation becomes
\begin{equation}
    \frac{q^{2}c^{2}}{v_\mathrm{A}^{2}z}=1+\sum_{\beta}\frac{\eta_{\beta}^{2}}{\sqrt{2}qu_{\beta}z}Z(\hat{\zeta}_{\beta}^{s}),
    \label{eq:disp_rel_adim}
\end{equation}
where
\begin{equation}
    \hat{\zeta}_{\beta}^{s}=\frac{z+sr_{\beta}+\mathrm{i}\tilde{\nu}_{\beta}}{\sqrt{2}qu_{\beta}}.
\end{equation}

\section{Numerical Results}
\label{sec:Numerical-Results}

In order to numerically solve the dispersion relation we consider a plasma composed of electrons, protons and dust particles with the following parameters: $B_{0}=1$\,G, $n_{i0}=10^{9}$\,cm$^{-3}$, $T_{i}=10^4$\,K, $T_{e}=T_{i}$ and $a=10^{-4}$\,cm, unless told otherwise. These are the same parameters employed in the literature to study Alfvén waves in a dusty plasma using a kinetic approach \citep[see e.g.][]{Juli_2005,Ziebell_2005,Gaelzer_2008,Gaelzer_2010} and are typical of stellar winds coming from carbon-rich stars \citep{tsytovich_2004}. 

This choice of parameters is motivated by the fact that Alfvén waves have been proposed as a possible accelerating mechanism for stellar winds \citep[see e.g.][]{alazraki_1971solarwind,Falceta_Goncalves_2002}, and carbon-rich stars are notable for losing great amounts of their mass by way of powerful stellar winds \citep{Knapp1987_MassLoss,Lafon1991_MassLoss}. Likewise, the Sun loses some of its mass by way of its wind, which can also be accelerated by damping of Alfvén waves, but we point out that the solar wind may present plasma temperatures of about $100$\,eV in the vicinity of the Sun's atmosphere \citep{kohnlein1996radial}, much larger than those of carbon-rich stars (around $1$\,eV)\@. In the range of temperatures typical to the solar corona, secondary electron emission gains importance as another charging mechanism, which is not considered in our theory.

Moreover, since the sun is an oxygen-rich star, it is more likely the formation of silicate particles in its surroundings, as opposed to carbon-rich stars where carbon particles will be more common \citep{nanni2021dust}\@. Silicate composed grains have a smaller photoelectric yield \citep{Feuerbacher_1972}, which will reduce the current of photoelectrons emitted from these particles when compared to carbon grains. Furthermore, silicate grains may be considered as insulators which present a higher secondary emission yield \citep{chow1993role}.

Near the Sun, carbon grains sublimate around $4\,\text{R}_{\odot}$ while silicate grains can exist up to $2\,\text{R}_{\odot}$ \citep{mann2004dust}. On the other hand, carbon stars are generally cooler and may feature a circumstellar dust envelope formed mainly by carbon grains \citep{wallerstein_1998_carbonstars}. 

Therefore, although the Sun has a higher surface temperature (and radiation flux) than most of carbon-rich stars, the discussion above makes us believe that the photoemission process will be more significant in relation to other charging mechanisms in a carbon-rich star. And since our work is more focused in the effects that arise when the photoionization process is included in the theory, we believe that the set of parameters typical of carbon-rich stars are better suited for our purposes. 

Regarding the parameters related to the photoelectric current, we consider that the dust particles are in a distance of $r_\mathrm{d}=2\,r_\mathrm{s}$ from the star and are exposed to a black body radiation spectrum given by equation~\eqref{eq:blackbody}, that emanates from a stellar surface with temperature $T_\mathrm{s}$ between $4500$ and $5000$\,K. This range of temperatures is within the observed values for some of the hotter carbon stars \citep{wallerstein_1998_carbonstars}, and is chosen with the purpose of analysing several situations where the equilibrium dust charge assumes negative, neutral and positive values.

We consider that the dust grains are composed by carbon with work function $\phi=4.6$\,eV, a value accepted for both graphite and glassy carbon \citep{Sohda_1997_glassycarbon}, maximum photoelectric efficiency $\chi_\mathrm{m}=0.05$ \citep{Feuerbacher_1972} and dust temperature $T_\mathrm{d}=300$\,K, which is within the range of temperatures of dust in the inner circumstellar dust shells of carbon stars \citep{gail_sedlmayr_2014}. This set of parameters are related only to the photoelectric current and are maintained fixed, given that the variation of this current is achieved by varying the stellar surface temperature.

It is worth mentioning that some of the chosen parameters may vary significantly in a stellar environment. For example, instead of having a constant value for all grains, the dust sizes are often described by a distribution function where they can assume a range of values from about $10^{-4}$ to $10^{-7}$\,cm \citep{kruger1997two}, so that the value of $a=10^{-4}$\,cm used in this work could be considered as a superior limit. A different grain size should modify the charging currents, changing its equilibrium electric charge, and the inelastic collision frequencies given by equation~\eqref{eq:colision_freq}. In future works, we intend to include an analysis of the consequences that a distribution of grain sizes brings to the propagation and damping of the waves.

Moreover, the densities and temperatures of plasma species will change the absorption rates of the plasma particles by the dust grains. To assume a similar temperature for ions and electrons could be questionable, since the electron population gain energy from emitted photoelectrons, raising its temperature above the ion temperature. On the other hand, ions should have smaller temperature, in the range of $1$--$2\times10^{3}$\,K, in order to allow the condensation and formation of dust grains in the vicinity of the star surface \citep{gail1984dust,gail_sedlmayr_2014}.

We also point out that the chosen ambient magnetic field value will not modify qualitatively the obtained results, given that the absorption currents are derived from the OML theory, which neglects the presence of the magnetic field in the collision trajectories between plasma particles and dust grains. Thereby, this parameter will only modify the cyclotron frequency and Alfvén velocity, changing only the values of normalized quantities. However, as mentioned in Section~\ref{subsec:Dust-Charging-Processes}, a sufficiently strong magnetic field, where the electron Larmor radius is comparable or smaller than the dust particle radius, could modify significantly the absorption of plasma particles by the dust grains \citep{salimullah2003dust,lange_2016,Kodanova+2019}.

Even though we will investigate some consequences of plasma temperature modification, we focus our analysis mainly in the changes that the photoionization process brings to the waves' properties, utilising the fixed set of parameters mentioned at the beginning of this section.

As pointed out before, the addition of the photoelectric current will result in a different equilibrium dust charge which may modify the dispersion relation through the quasi-neutrality condition. Therefore, we start our investigation by studying this modification of the equilibrium dust charge number $Z_\mathrm{d}$ and electron density $n_{e0}$ for different values of dust density, obtained from $\varepsilon=n_{d0}/n_{i0}$.

\begin{figure}
    \centering
    \begin{minipage}[c]{\columnwidth}
        \centering
        \includegraphics[width=\columnwidth]{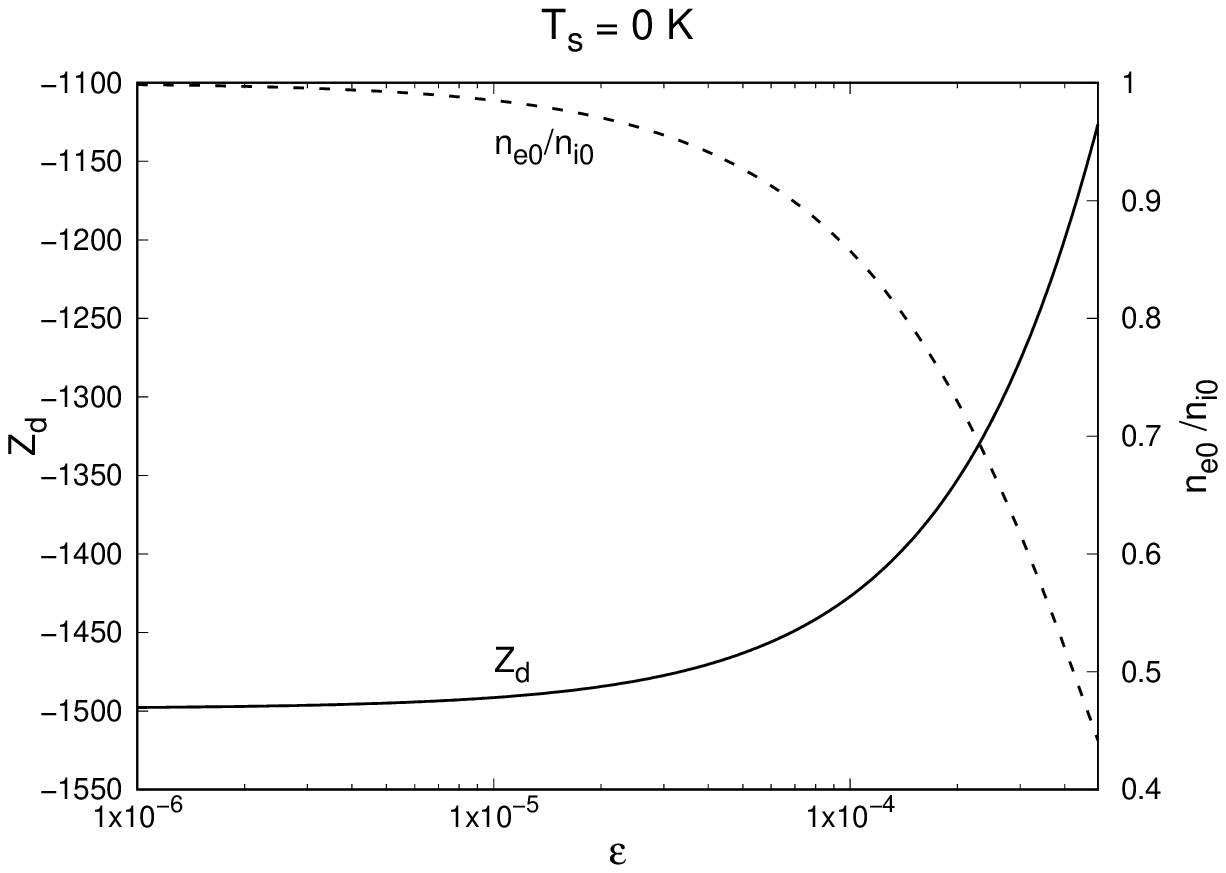}
        \vspace{\floatsep}
    \end{minipage} 
    \begin{minipage}[c]{\columnwidth}
        \centering
        \includegraphics[width=\columnwidth]{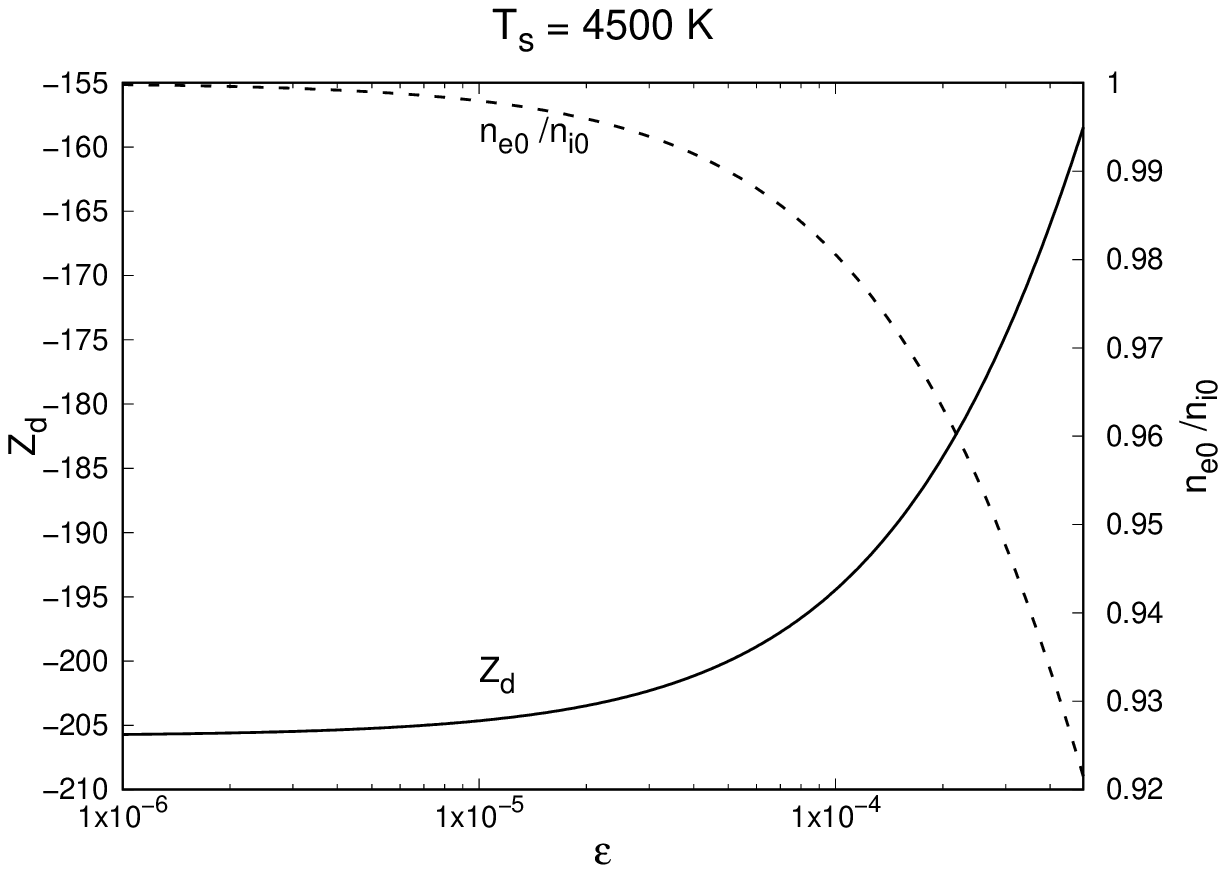}
        \vspace{\floatsep}
    \end{minipage} 
    \begin{minipage}[c]{\columnwidth}
        \centering
        \includegraphics[width=\columnwidth]{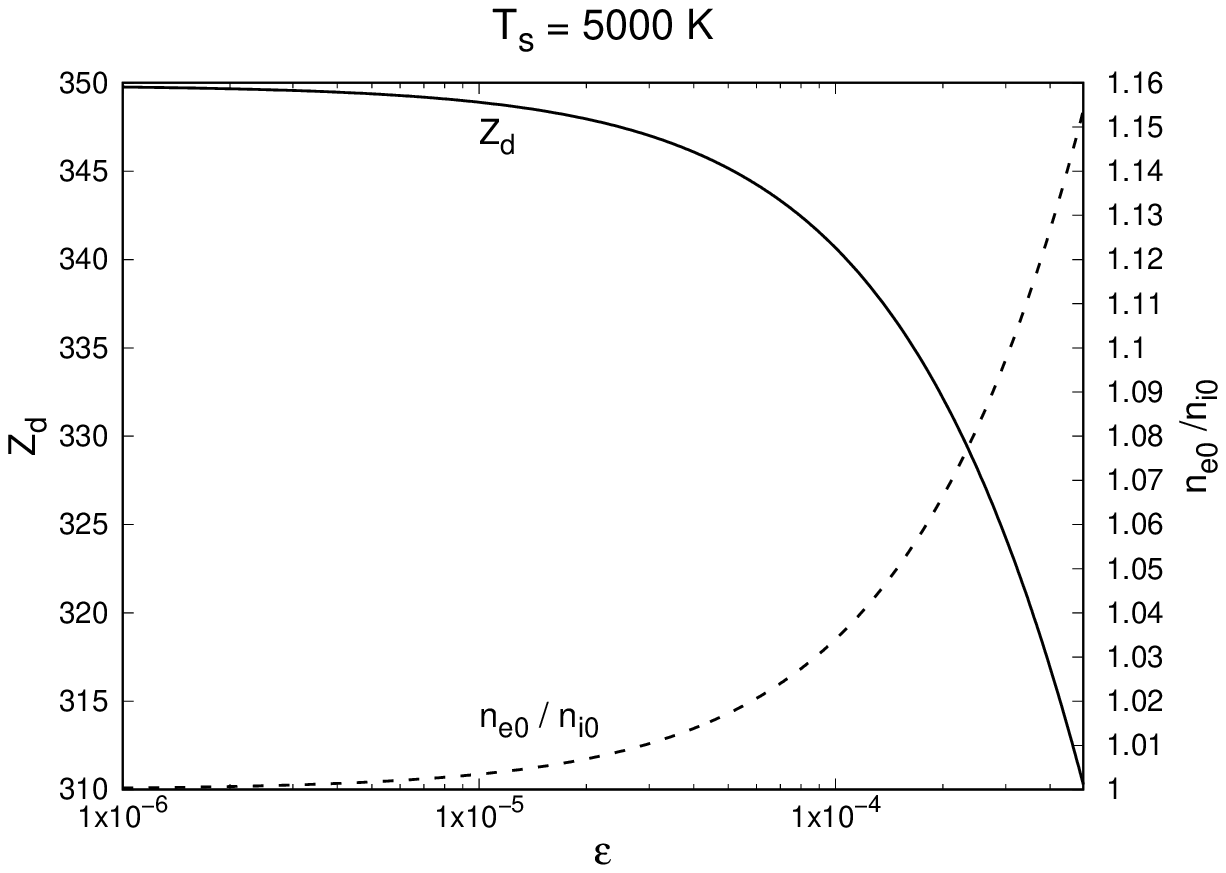}
    \end{minipage}
    \caption{Equilibrium dust charge number $Z_\mathrm{d}$ and electron density fraction $n_{e0}/n_{i0}$ as function of dust density fraction $\varepsilon=n_{d0}/n_{i0}$ for: only absorption of plasma particles as charging mechanism, i.e., no incidence of radiation (top panel); adding photoionization as charging mechanism with $T_\mathrm{s}=4500$\,K (mid panel) and $T_\mathrm{s}=5000$\,K (bottom panel). The considered parameters are $T_{i}=10^4$\,K, $T_{e}=T_{i}$, $n_{i0}=10^{9}$\,cm$^{-3}$ and $a=10^{-4}$\,cm.}
    \label{fig:eq_charge}
\end{figure}

The top panel of Fig.~\ref{fig:eq_charge} shows the case where only the absorption current is taken into account, this situation is symbolised with a surface stellar temperature of $T_\mathrm{s}=0$\,K, i.e., there is no incident radiation on the dust particles. For the used parameters, we notice that the equilibrium dust charge is always negative for all values of $\varepsilon$. This is expected since the electron absorption rate is greater than the ion absorption rate. Consequently, the equilibrium electron density is always smaller than the ion density.

The mid and bottom panels of Fig.~\ref{fig:eq_charge} include both the absorption and photoionization currents, with stellar temperatures of $T_\mathrm{s}=4500$\,K and $T_\mathrm{s}=5000$\,K, respectively. These two temperature values represent distinct cases, the former keeping the dust charge still negative, but with smaller absolute value, while the latter is a case where the dust particles acquire a positive electrical charge and, consequently, the electron density becomes larger than the ion density.

\begin{figure}
    \centering
    \includegraphics[width=\columnwidth]{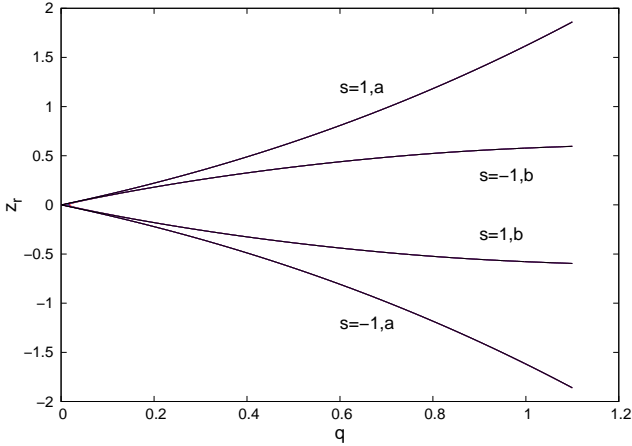}
    \caption{Real part of the normalized frequency $z_\mathrm{r}$ for the roots obtained using $s=\pm1$ as a function of $q$ for $\varepsilon=0.0$, $\varepsilon=1.25\times10^{-6}$, $\varepsilon=2.5\times10^{-6}$, $\varepsilon=3.75\times10^{-6}$ and $\varepsilon=5.0\times10^{-6}$. Parameters used are $B_{0}=1$\,G, $n_{i0}=10^{9}$\,cm$^{-3}$, $T_{i}=10^4$\,K, $T_{e}=T_{i}$, $a=10^{-4}$\,cm and $T_\mathrm{s}=4500$\,K.}
    \label{fig:zr_q}
\end{figure}

To start investigating the modification of the photoemission process in the dispersion relation we plot the real part of the two roots obtained from equation~\eqref{eq:disp_rel_adim} for each of the signs $s=\pm1$, using the same set of parameters as in Fig.~\ref{fig:eq_charge} with $T_\mathrm{s}=4500$\,K. Fig.~\ref{fig:zr_q} shows $z_\mathrm{r}(q)$, the real part of the roots of equation~\eqref{eq:disp_rel_adim}, as functions of $q$. 

For each wave mode and each value of $q$, there are two solutions of Eq.~\eqref{eq:disp_rel_adim}, each with either positive or negative values for $z_r$, but both with the same value for $z_i$\@. In Fig.~\ref{fig:zr_q}, we show dispersion relation curves for wave modes with either positive or negative frequencies.  The positive solutions correspond to forward-propagating waves (phase velocity $v_\phi= \omega_r/k >0$), relative to the ambient magnetic field, whereas the negative solutions are related to backward-propagating waves $(v_\phi < 0)$\@. The solutions show that the dispersive characteristics of a given parallel-propagating mode does not depend on the propagation direction, as is expected due to the symmetries of the plasma, the dust grains and the charging mechanisms.

A similar plot has already appeared in \citet{Juli_2005}, where the dispersion relation was evaluated for the same parameters without the photoelectric current. The resemblance between these figures could lead one to believe that the influence of the photoionization mechanism is negligible, but this is due to the relatively extended range of frequencies and wavenumbers shown in Fig.~\ref{fig:zr_q}\@.  The effect of the photoionization appears on the small-wavenumber scale.

Nevertheless, this plot is useful to identify the wavemodes predicted by the dispersion relation, which are recognized from well known textbooks \citep[see e.g.][]{krall1973principles}. The roots identified as $s=1,a$ and $s=-1,a$ correspond to the whistler branch whereas the roots $s=-1,b$ and $s=1,b$ are the ion cyclotron modes. 
The wave modes can be identified by their distinct circular polarizations. The whistler mode is the solution of equation (\ref{eq:Dispersion-equation}) with the ratio of the amplitudes of the electric field $E_1/E_2 = -i$, corresponding to a right-handed $(R)$ circularly-polarized electromagnetic wave propagating parallel to the ambient magnetic field.  On the other hand, for the ion-cyclotron mode, $E_1/E_2= +i$, corresponding to left-handed $(L)$ circularly-polarized waves.

A simple rôle of thumb can also be employed to distinguish the transverse modes.  Since the effect of the dust becomes negligible in the high wavenumber (small wavelength) limit, the asymptotic properties of the modes depicted in Fig.~\ref{fig:zr_q} are identical to the modes propagating in a dust-free plasma.  Hence, since the whistler mode tends to $\omega = |\Omega_e|$ (for $q\to\infty$), whereas the ion-cyclotron behaves as $\omega \to \Omega_i$, the dispersion curves in Fig.~\ref{fig:zr_q} with $|z_r| > 1$ are immediately identified as the whistler mode.

Both modes couple for small values of $q$ becoming the Alfvén mode. Each of the curves actually corresponds to the superposition of five curves, obtained with $\varepsilon=0.0$, $\varepsilon=1.25\times10^{-6}$, $\varepsilon=2.5\times10^{-6}$, $\varepsilon=3.75\times10^{-6}$ and $\varepsilon=5.0\times10^{-6}$.

\begin{figure}
    \centering
    \begin{minipage}[c]{\columnwidth}
        \centering
        \includegraphics[width=\columnwidth]{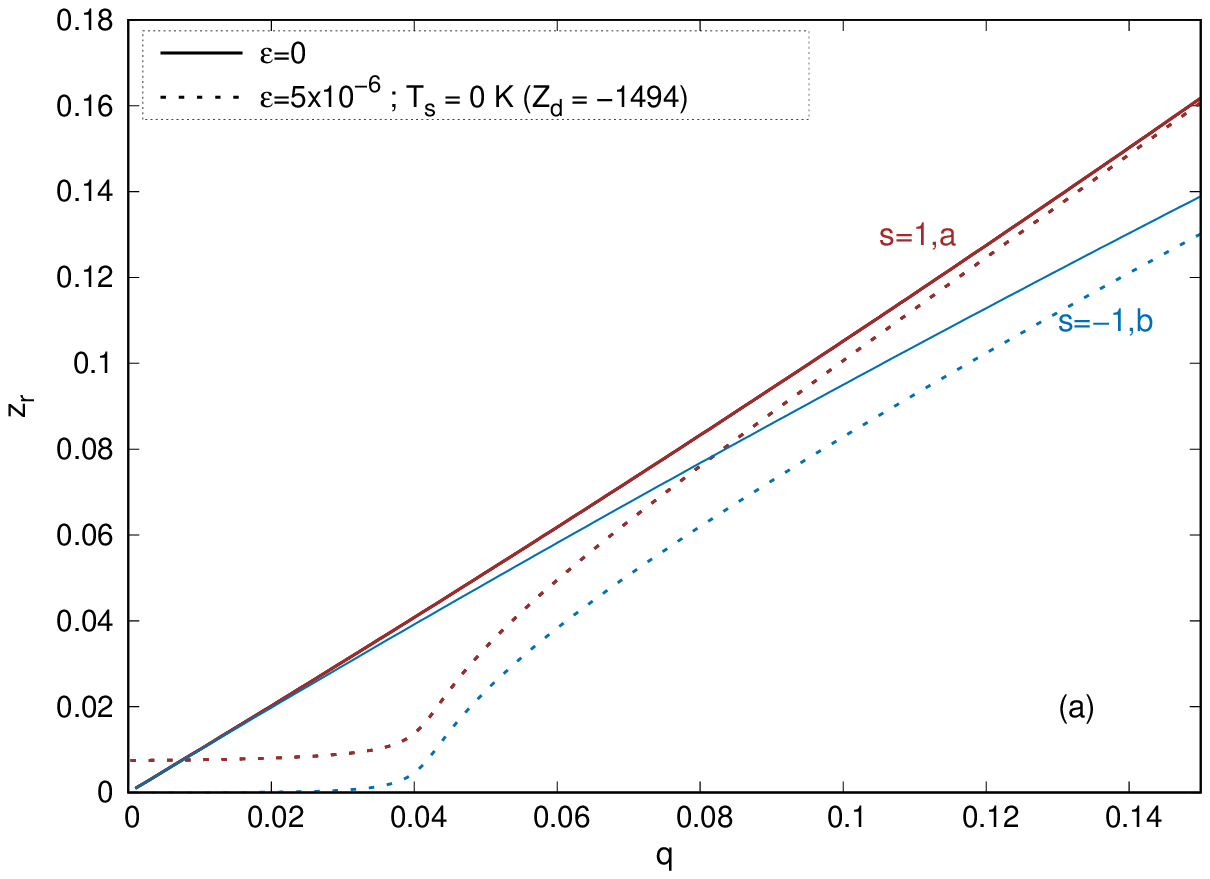}%
    \end{minipage}\vspace{\floatsep}
    \begin{minipage}[c]{\columnwidth}
        \centering
        \includegraphics[width=\columnwidth]{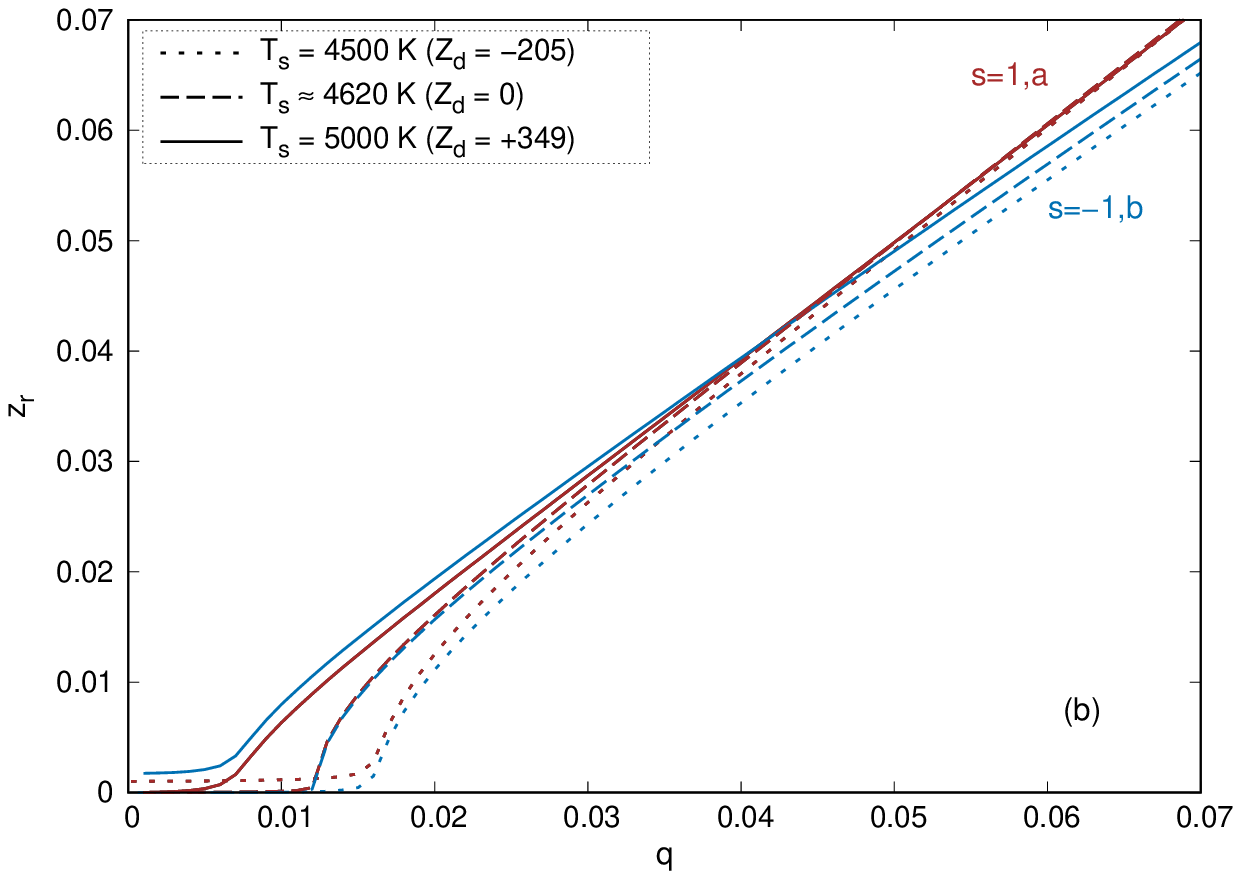}
    \end{minipage}
    \caption{Plot of $z_\mathrm{r}$ as a function of $q$ for: (a) a dustless plasma (continuous lines) and $\varepsilon=5.0\times10^{-6}$ considering only the absorption current (dotted lines); (b) $\varepsilon=5.0\times10^{-6}$ considering both the absorption and photoelectric currents, with stellar temperature of $T_\mathrm{s}=4500$\,K (dotted lines), $T_\mathrm{s}\simeq4620$\,K (dashed lines) and $T_\mathrm{s}=5000$\,K (continuous lines). Brown lines ($s=1,a$) are the whistler mode and blue lines ($s=-1,b$) are the ion cyclotron mode.}
    \label{fig:zr_q_small}
\end{figure}

Fig.~\ref{fig:zr_q_small} shows the real part of the frequencies $z_\mathrm{r}$ for small values of $q$, where is possible to see the modification caused by the presence of dust. The whistler branch is shown in brown colour ($s=1,a$) while the ion cyclotron branch is in blue ($s=-1,b$).

We see in Fig.~\ref{fig:zr_q_small}(a) two cases already known in the literature, the case of a dustless plasma, given by the continuous lines, and the case where dust grains are present and they are charged only by absorption of plasma particles, given by the dotted lines. For $\varepsilon=0$, it is possible to observe the coupling between the whistler and ion cyclotron modes into the Alfvén mode for small $q$. On the other hand, the presence of a small density of $\varepsilon=5.0\times10^{-6}$ of dust negatively charged results in a detachment between the modes for all values of wavenumber. We also note that both modes show a region with null group velocity (constant $z_\mathrm{r}$), a feature observed only in the presence of dust.

Fig.~\ref{fig:zr_q_small}(b) shows some cases where there is incidence of radiation on the surface of dust particles, resulting in an equilibrium dust charge with negative (dotted lines), null (dashed lines) or positive (continuous lines) values. We see that, in the cases where the grains assume zero or positive values, the lines corresponding to the whistler and ion cyclotron modes meet in a certain region of wavenumber values. That is, the detachment for all values of $q$ between these two modes occurs only in the case of negatively charged dust particles.

We also observe that the presence of the photoionization process reduces the interval where the modes present constant values of $z_\mathrm{r}$, and that the separation of the lines in the region of null group velocity reduces for smaller values of $|Z_\mathrm{d}|$. This feature was also remarked by \citet{Gaelzer_2010} and is related to the charge imbalance between electrons and ions which occurs in the presence of dust particles. Since this imbalance is attenuated for smaller values of $|Z_\mathrm{d}|$, the separation between the curves is also attenuated and, as a consequence, for positively charged dust grains, where the electron population is greater than the ion population, the ion cyclotron mode achieves bigger values of $z_\mathrm{r}$ than the whistler mode for small values of $q$. 

\begin{figure}
    \centering
    \begin{minipage}[c]{\columnwidth}
        \centering
        \includegraphics[width=\columnwidth]{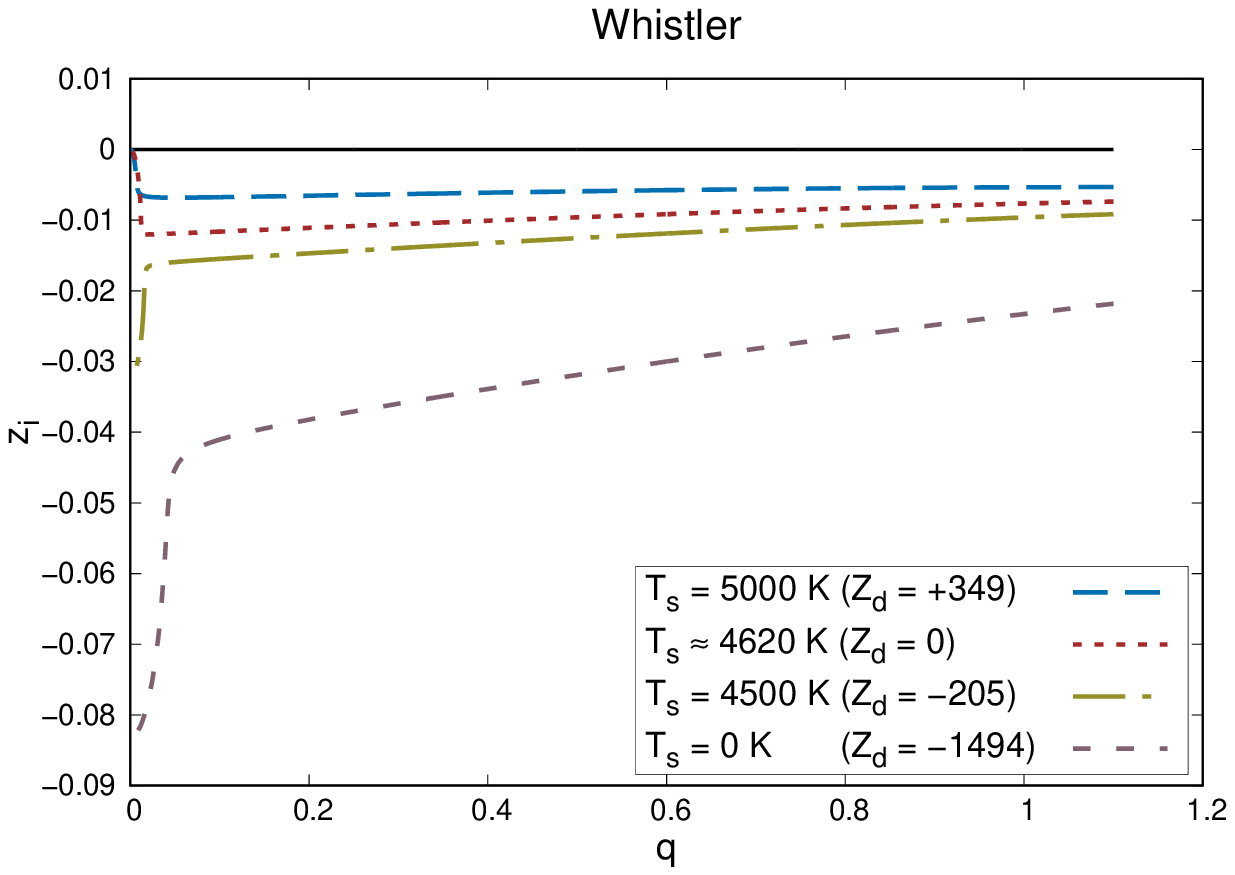}%
    \end{minipage}\vspace{\floatsep}
    \begin{minipage}[c]{\columnwidth}
        \centering
        \includegraphics[width=\columnwidth]{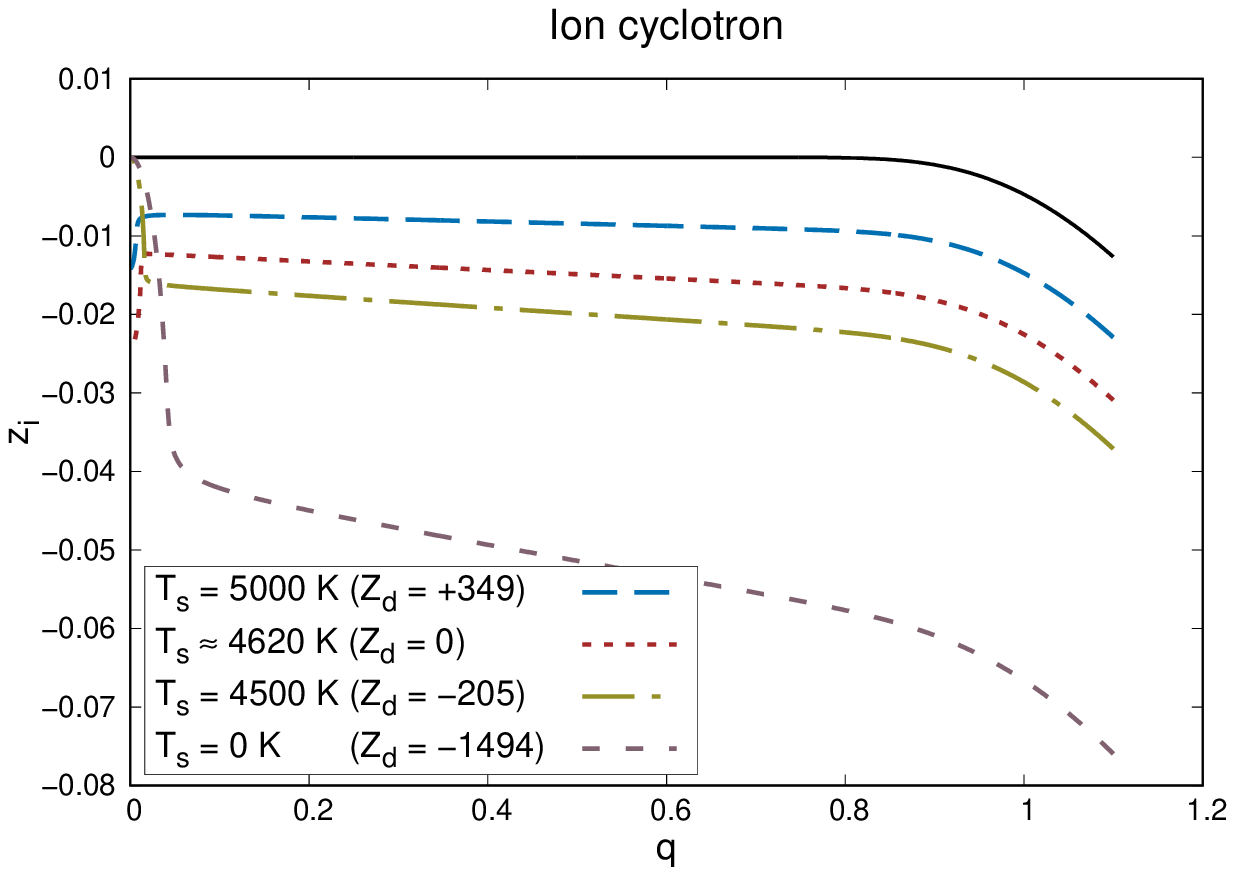}
    \end{minipage}
    \caption{Imaginary part of the normalized frequency $z_\mathrm{i}$ as a function of $q$ for: the roots labelled as $s=1,a$ and $s=-1,a$ in Fig.~\ref{fig:zr_q}, i.e., the whistler branch (top panel); the roots labelled as $s=1,b$ and $s=-1,b$ in Fig.~\ref{fig:zr_q}, i.e., the ion cyclotron branch (bottom panel). Continuous lines are the cases of a dustless plasma ($\varepsilon=0$) while discontinuous lines represent the cases with $\varepsilon=5\times10^{-6}$ for several values of $T_\mathrm{s}$.}
    \label{fig:zi_q}
\end{figure}

Fig.~\ref{fig:zi_q} shows the corresponding imaginary parts $z_\mathrm{i}$ of the roots depicted in Fig.~\ref{fig:zr_q} as a function of $q$ for the whistler branch ($s=1,a$ and $s=-1,a$) and ion cyclotron branch ($s=1,b$ and $s=-1,b$). In these figures we show the case of a dustless plasma (continuous lines) and the case of a plasma with the presence of a small density $\varepsilon=5\times10^{-6}$ of dust particles (discontinuous lines) exposed to different values of radiation flux, determined by the stellar surface temperature $T_\mathrm{s}$.

First of all, we notice that $z_i(q)$ depends only on the wave mode and plasma parameters and not on the propagation direction.  Again, this is expected due to the general symmetry of the system.  Since $z_i$ is proportional to the absorption coefficient of the waves, which is due to the irreversible processes of wave-particle interactions that occur in the plasma, these processes are the same for both forward- and backward-propagating waves in the system under consideration.

We see that in the absence of dust the whistler mode has negligible damping values for all values of $q$. For the ion cyclotron mode, the damping rate is also negligible for $\varepsilon=0$ in the region of small $q$, but becomes significant when $q\simeq1$, this is known as the ion cyclotron damping and is due to wave-particle resonance. We notice from the discontinuous lines that the presence of dust particles causes the appearance of a new damping mechanism in both modes, which tends to dominate over the ion cyclotron damping.

This new damping mechanism is already known \citep[see e.g.][]{Juli_2005,Ziebell_2005} and is caused by the imaginary term $\mathrm{i}\tilde{\nu}_{\beta}$ that appears in the dispersion relation in the presence of dust. This term corresponds to the average inelastic collision frequency between dust grains and plasma particles. \citet{Juli_2005} showed that in the approximation $\tilde{\nu}_{\beta}=0$ this new damping rate disappears, demonstrating that it is in fact due to the inelastic collisions that happen in the presence of dust.

We can see the effect of photoionization in this damping mechanism by comparing the different discontinuous lines in both panels of Fig.~\ref{fig:zi_q}, which represent distinct photoionization currents incident on the grain's surface, modifying the dust equilibrium electrical charge and, consequently, the inelastic collision frequency. We consider the case without photoionization ($T_\mathrm{s}=0$\,K) and three cases with the presence of radiation where the grains acquire a negative ($T_\mathrm{s}=4500$\,K), neutral ($T_\mathrm{s}\simeq4620$\,K) or positive ($T_\mathrm{s}=5000$\,K) equilibrium electrical charge.

It is possible to see that, in general, a larger radiation flux will diminish the damping rate of both modes for the considered parameters. However, in the region of small wavenumber the behaviour of the curves may change with the change of sign of the dust electrical charge. For negatively charged grains, whistler waves (right circularly-polarized) will be strongly absorbed for very small values of wavenumber, while the ion cyclotron waves (left circularly-polarized) present negligible damping rate in this region. This feature was already observed in works where only the absorption of particles is considered as charging mechanism of dust particles and, consequently, the grains can acquire only negative values of electrical charge \citep[see e.g.][]{Juli_2005,Ziebell_2005,Gaelzer_2010}.

Now, when photoionization is considered and when the intensity of radiation is big enough to allow the grains to acquire a neutral or positive electrical charge, we observe that the whistler waves will show negligible damping rate while the ion cyclotron waves will be strongly absorbed in the region of very small $q$.\@ 
Hence, we conclude that the polarization of the waves together with the dust electrical charge's sign will determine if the waves will show very high or low values of damping rate for small values of wavenumber.

\begin{figure}
    \centering
    \begin{minipage}[c]{\columnwidth}
        \centering
        \includegraphics[width=\columnwidth]{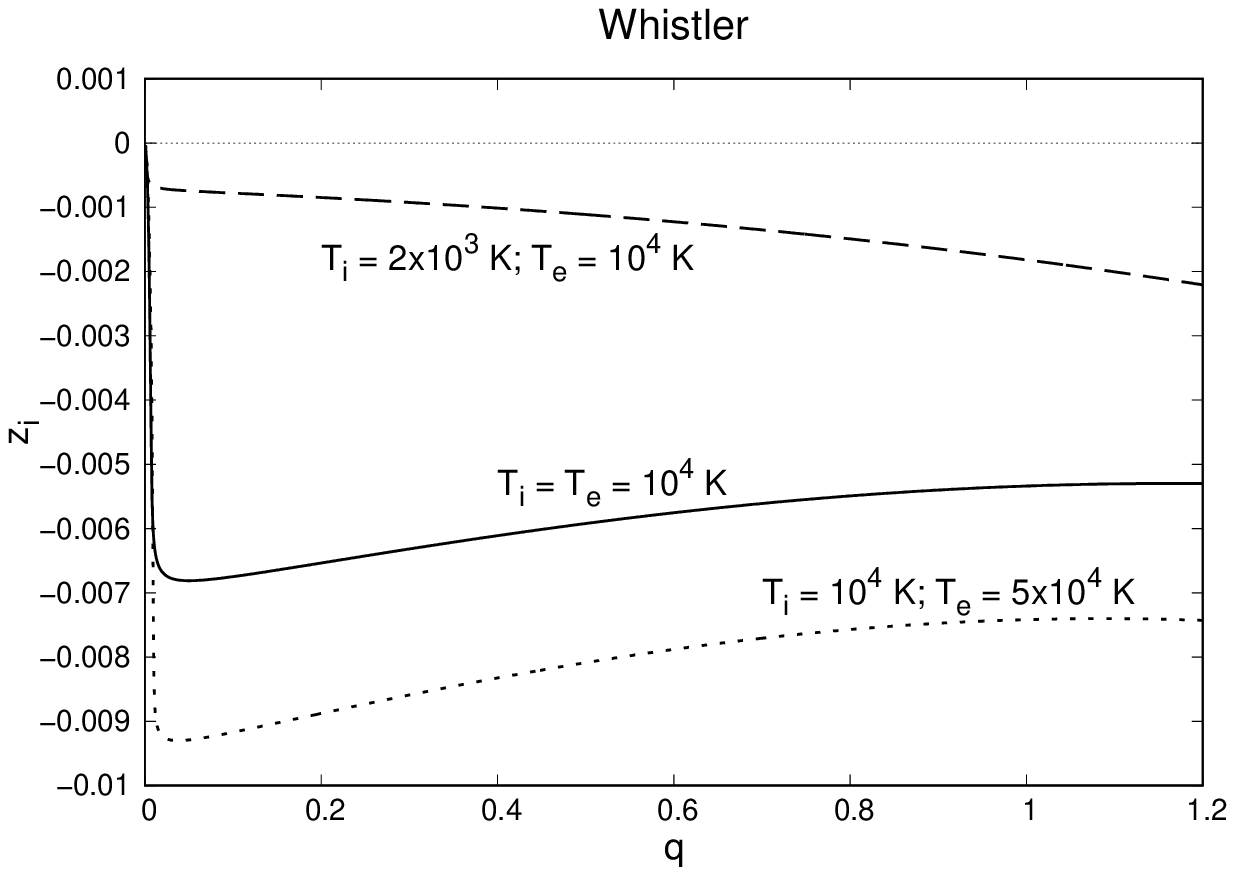}%
    \end{minipage}\vspace{\floatsep}
    \begin{minipage}[c]{\columnwidth}
        \centering
        \includegraphics[width=\columnwidth]{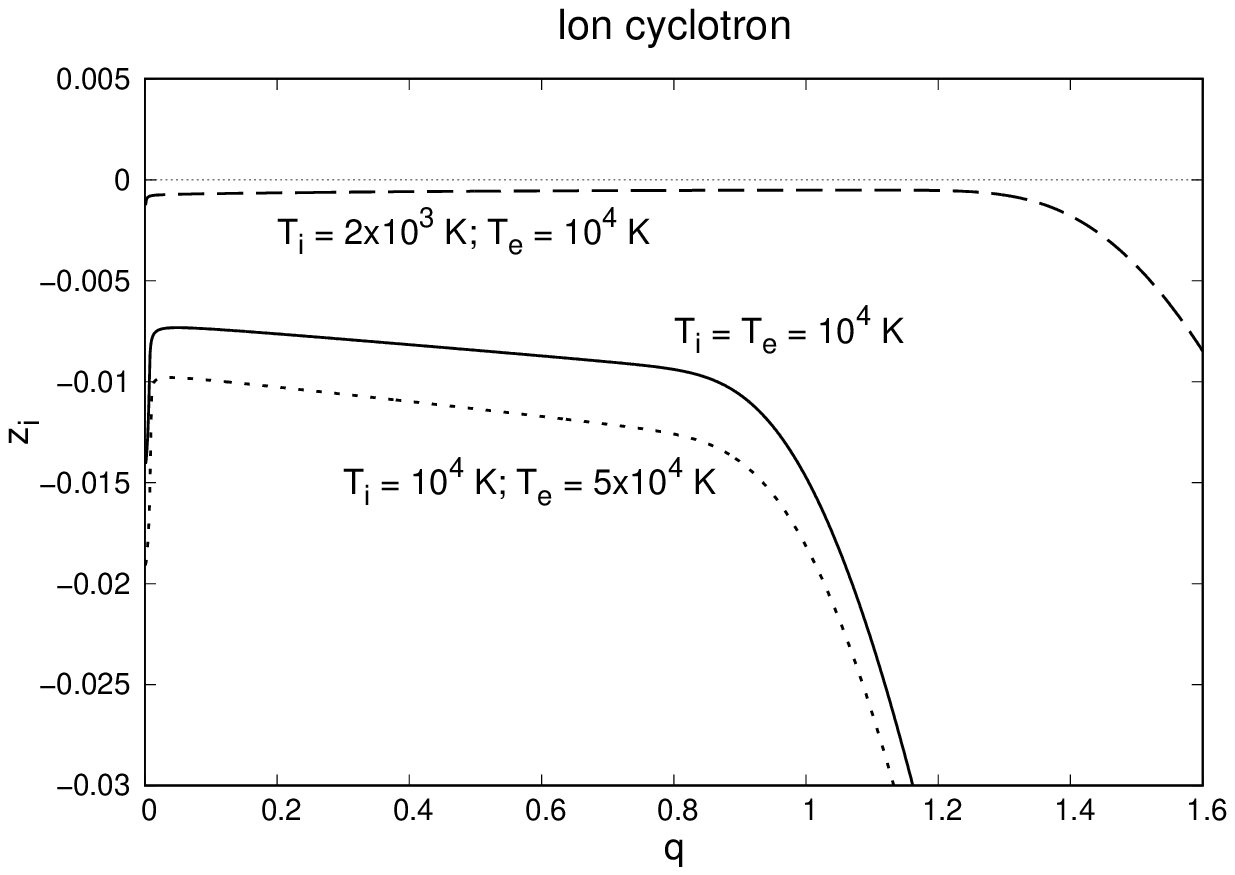}
    \end{minipage}
    \caption{Imaginary part of the normalized frequency $z_\mathrm{i}$ as a function of $q$ for: whistler waves (top panel); ion cyclotron waves (bottom panel). Different values of temperatures are considered: $T_{i}=T_{e}=10^{4}$\,K (continuous lines); $T_{i}=2\times10^{3}$\,K and $T_{e}=10^{4}$\,K (dashed lines); $T_{i}=10^{4}$\,K and $T_{e}=5\times10^{4}$\,K (dotted lines). Other parameters are the same as in Fig.~\ref{fig:zr_q}, with $T_\mathrm{s}=5000$\,K and $\varepsilon=5.0\times10^{-6}$.}
    \label{fig:zi_q_temp}
\end{figure}

The effects of plasma species' temperatures in the damping of whistler and ion cyclotron modes can be appreciated in Fig.~\ref{fig:zi_q_temp} where we consider the case of $T_{e}/T_{i}=5$. As mentioned before, this is a more realistic situation since the electron population will gain energy from photoelectrons emitted from dust particles. We compare this situation with the case $T_{e}=T_{i}$ (continuous lines) already showed in Fig.~\ref{fig:zi_q} for $T_\mathrm{s}=5000$\,K, this is done by raising the electron temperature to $T_{e}=5\times10^{4}$\,K (dotted lines) or by diminishing the ion temperature to $T_{i}=2\times10^{3}$\,K (dashed lines), maintaining the other specie's temperature in $10^{4}$\,K.

The panels in Fig.~\ref{fig:zi_q_temp} show that the increase in electron temperature will increase the damping rates of the modes, but it does not change qualitatively the dependence of $z_\mathrm{i}$ with $q$. However, when we change the ion temperature we observe a qualitative change in the curves together with the decrease in damping values. Now, the damping rate of the whistler mode rises with increasing wavenumber where before it would reduce for greater values of $q$ in the considered range. For the ion cyclotron waves we see that the damping rate has a large decrease when we reduce the ion temperature, with $z_\mathrm{i}$ being practically independent of $q$ until it reaches the region of the ion cyclotron damping, which now occurs for a greater value of wavenumber, around $q\simeq1.4$.

\begin{figure*}
    \centering
    \begin{minipage}[t]{0.49\textwidth}%
        \centering
        \includegraphics[width=1\columnwidth]{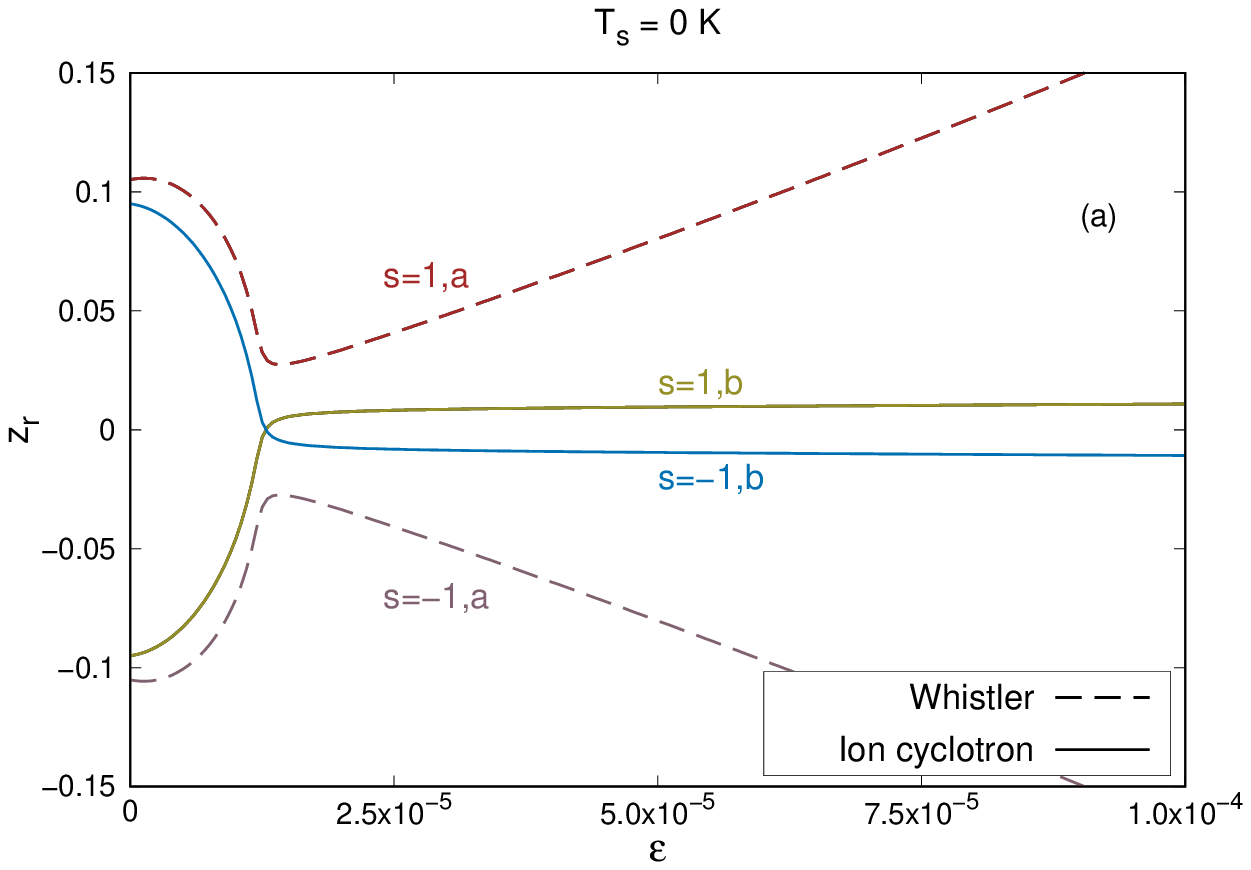}%
    \end{minipage}\hfill{}%
    \begin{minipage}[t]{0.49\textwidth}%
        \centering
        \includegraphics[width=1\columnwidth]{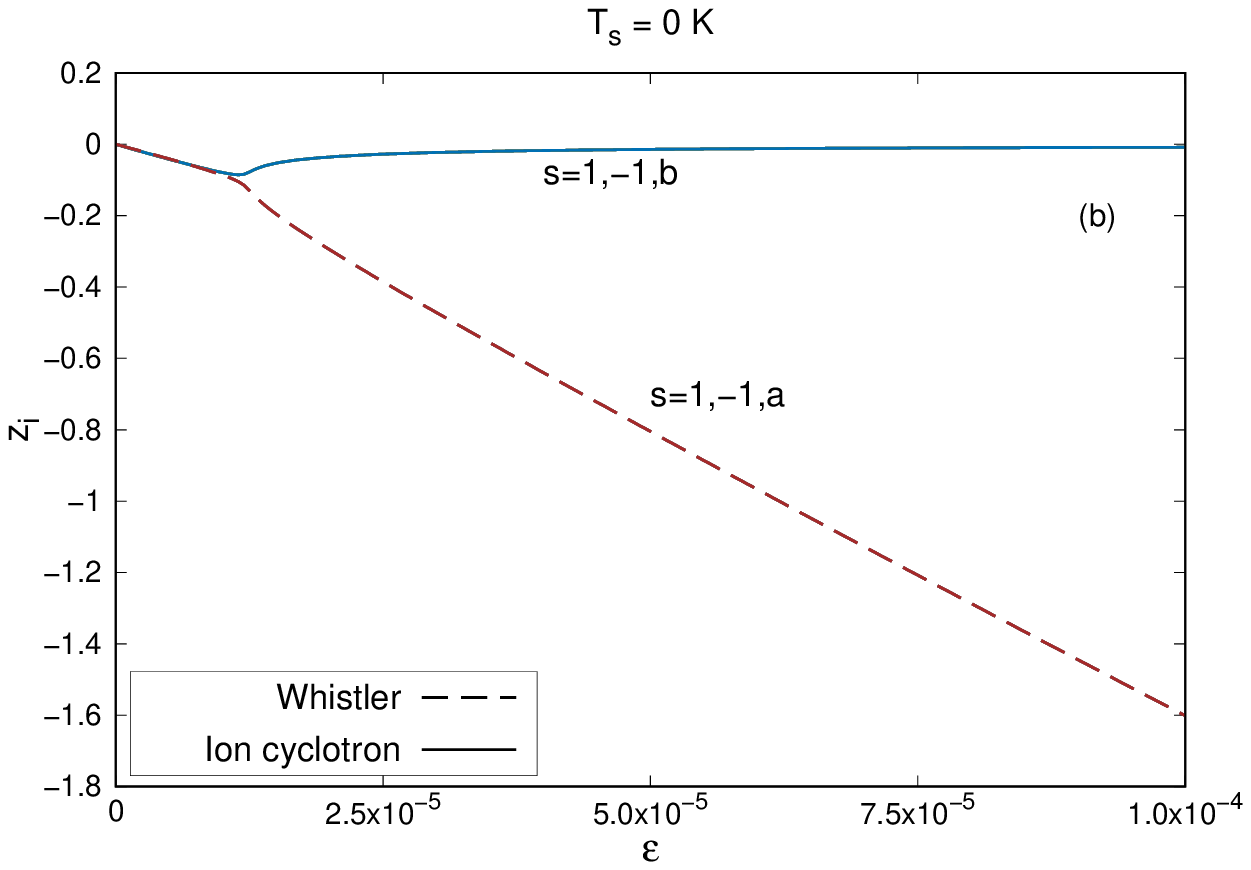}%
    \end{minipage}\vspace{\floatsep}
    \begin{minipage}[t]{0.49\textwidth}%
        \centering
        \includegraphics[width=1\columnwidth]{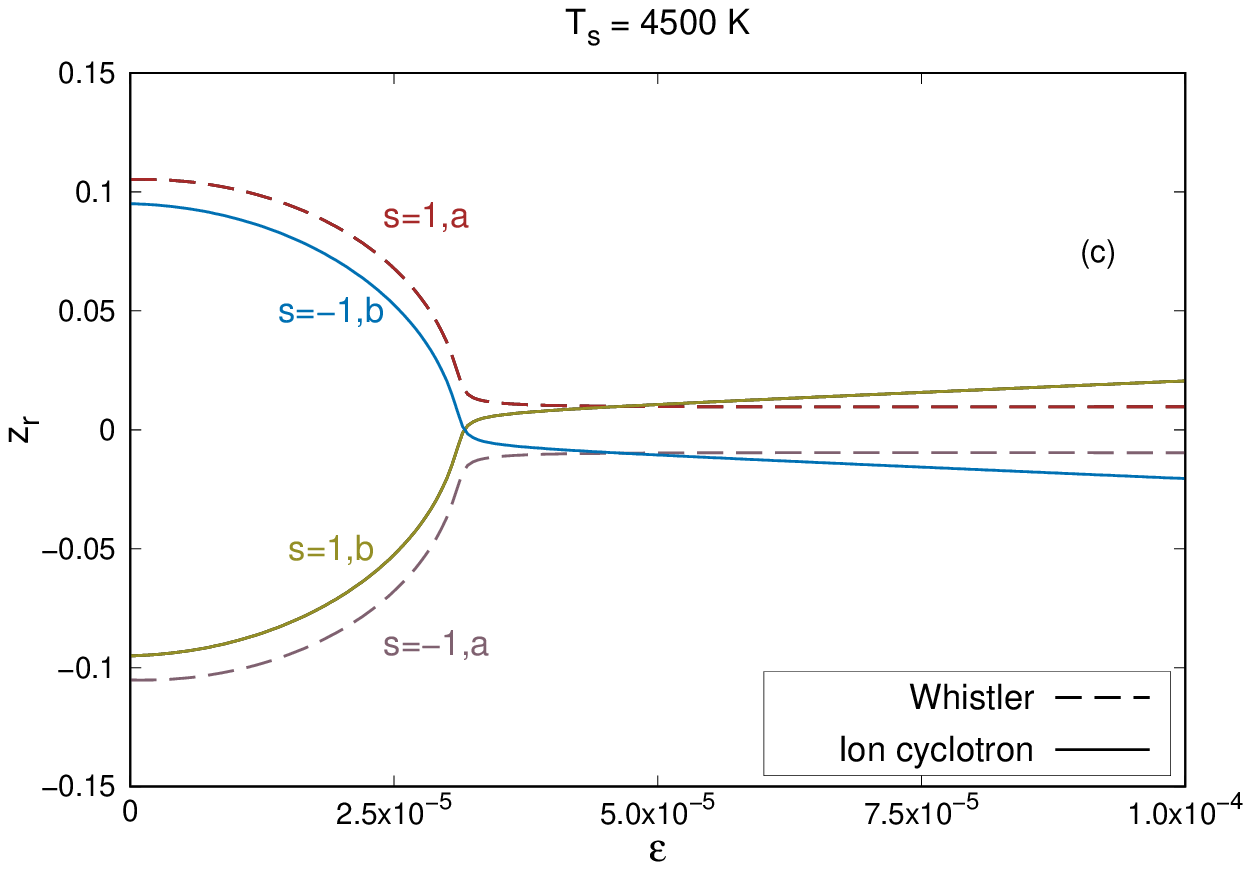}%
    \end{minipage}\hfill{}%
    \begin{minipage}[t]{0.49\textwidth}%
        \centering
        \includegraphics[width=1\columnwidth]{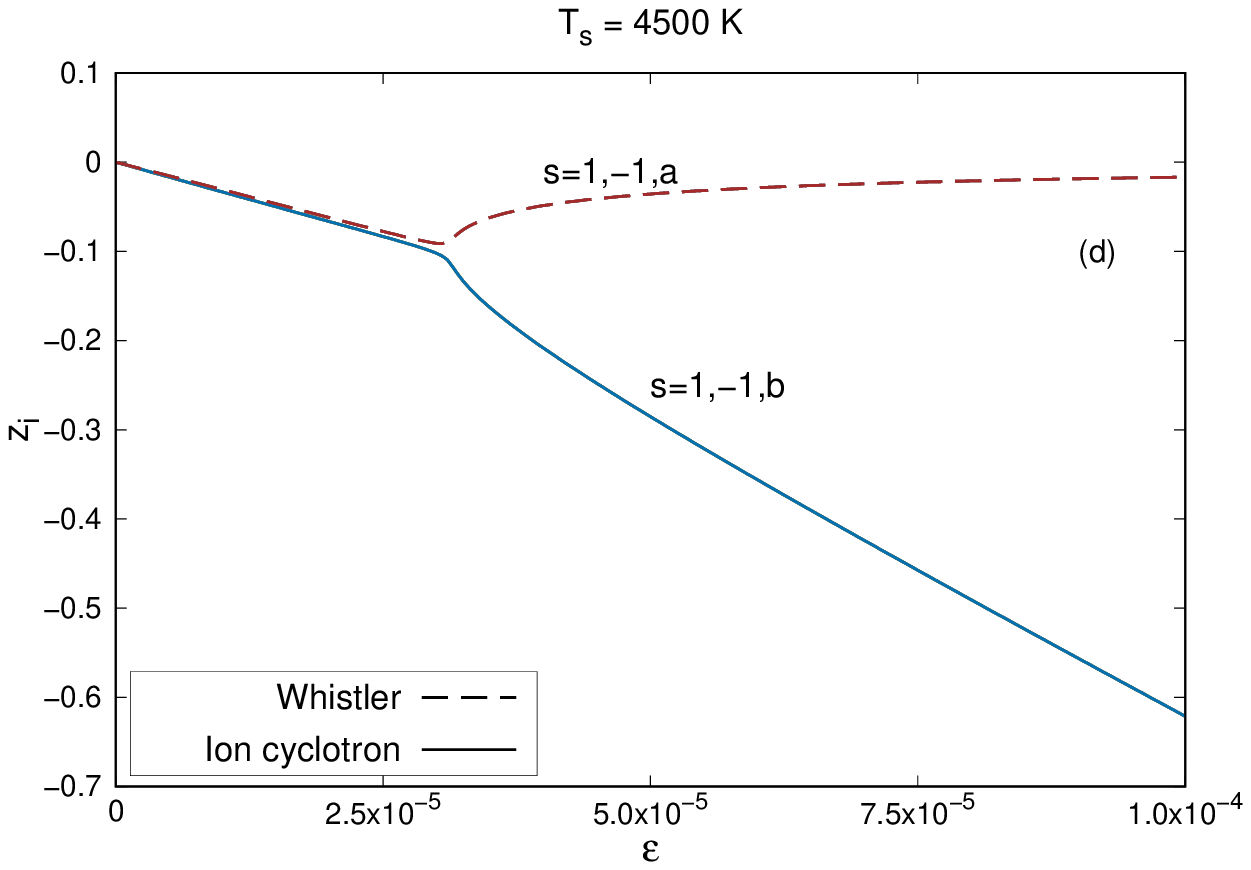}%
    \end{minipage}\vspace{\floatsep}
    \begin{minipage}[t]{0.49\textwidth}%
        \centering
        \includegraphics[width=1\columnwidth]{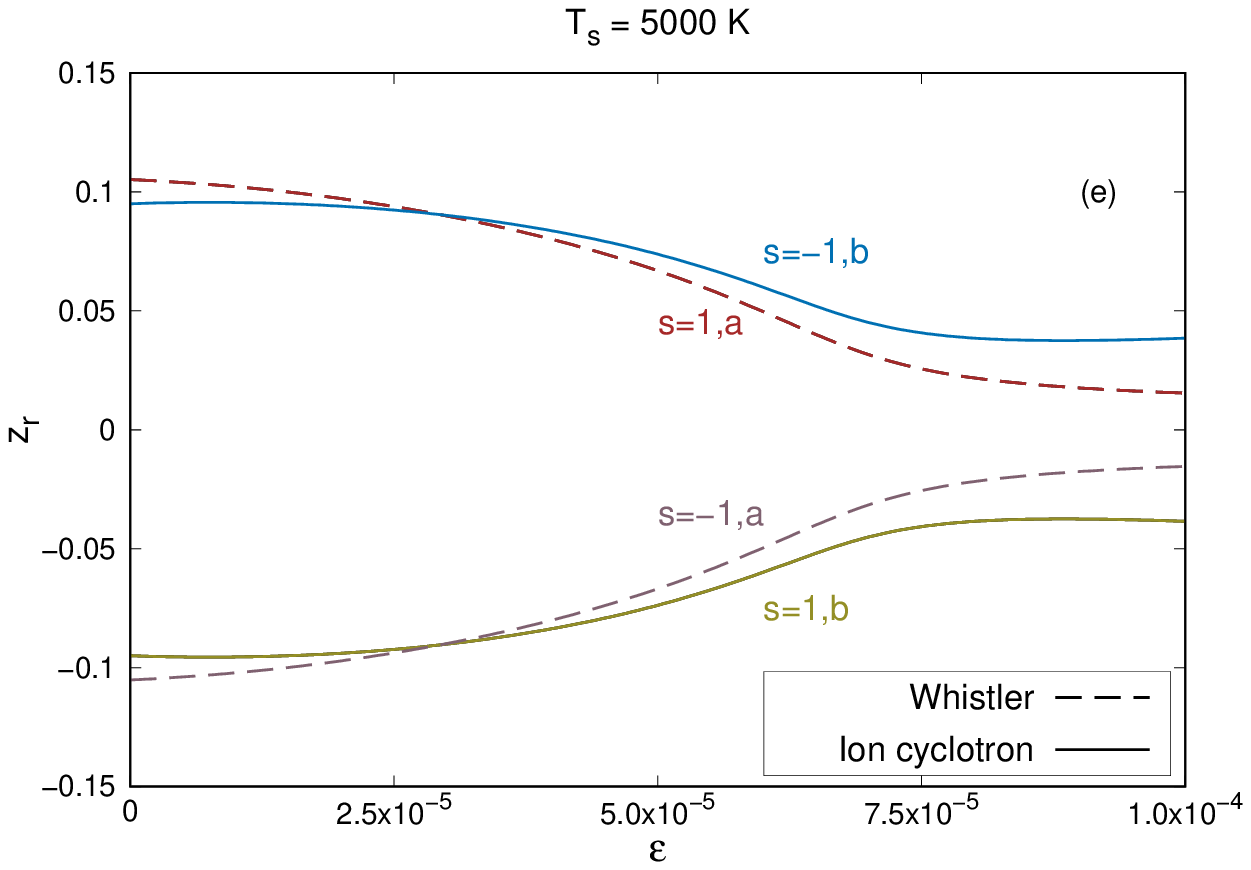}%
    \end{minipage}\hfill{}%
    \begin{minipage}[t]{0.49\textwidth}%
        \centering
        \includegraphics[width=1\columnwidth]{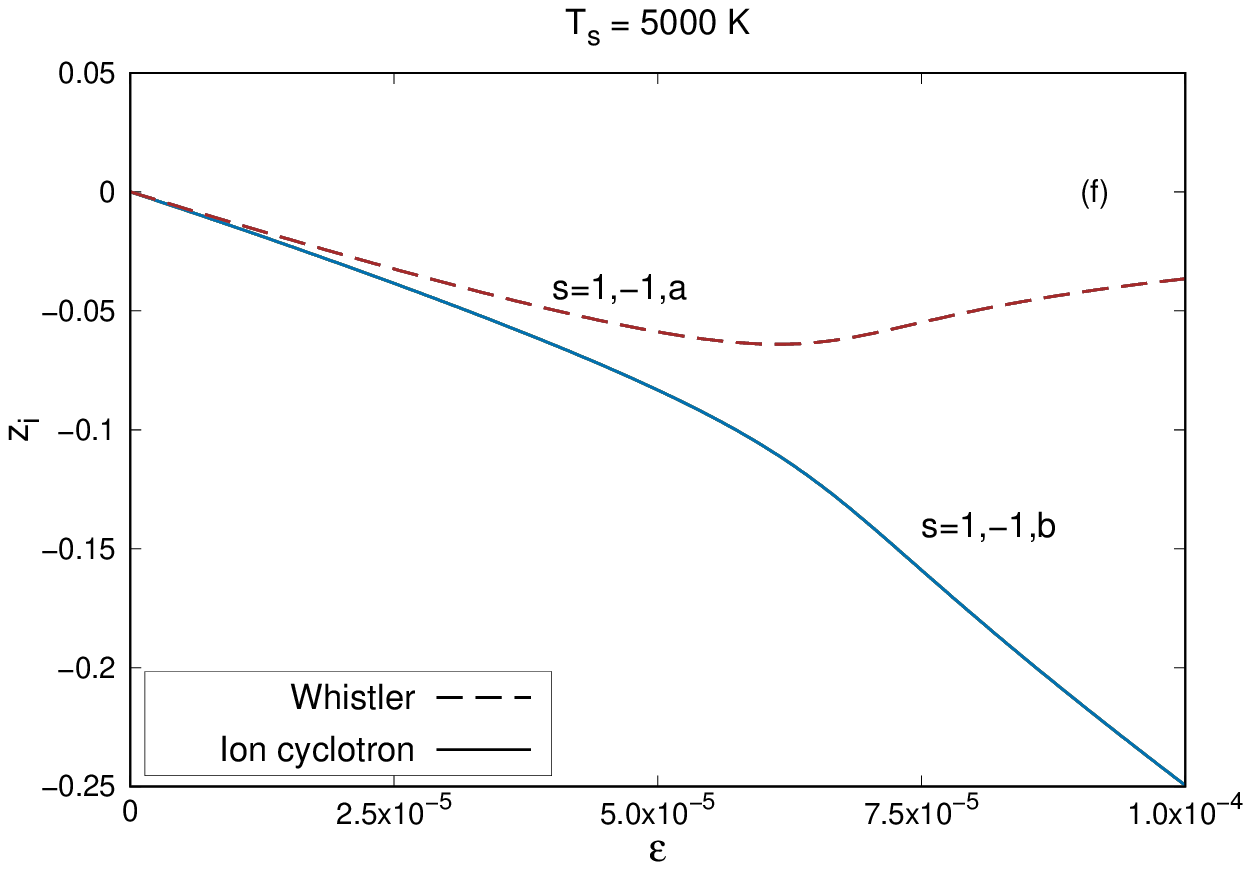}%
    \end{minipage}
    \caption{Real (left panels) and imaginary (right panels) parts of the normalized wave frequencies $z$ as a function of $\varepsilon$ with constant value of $q=0.1$ and three distinct values of radiation intensity, given by the stellar surface temperature $T_\mathrm{s}$. Other parameters are the same as in Fig.~\ref{fig:zr_q}.}
    \label{fig:z_eps}
\end{figure*}

Finally, we investigate the changes due to photoionization when we vary the dust density of the plasma for a fixed value of wavenumber. Fig.~\ref{fig:z_eps} shows the plots of the real and imaginary parts of the normalized wave frequency $z$ as a function of $\epsilon$ for a fixed value of $q=0.1$ and three values of stellar surface temperature $T_\mathrm{s}$. We identify the roots from Fig.~\ref{fig:zr_q} for $\epsilon=0$, the $s=1,a$ and $s=-1,a$ curves correspond to the whistler branch whereas the $s=-1,b$ and $s=1,b$ curves correspond to the ion cyclotron branch.

The top panels of Fig.~\ref{fig:z_eps} show the case without photoionization ($T_\mathrm{s}=0$\,K), which has already been studied within this formulation and the parameters used in this work \cite[see e.g.][]{Juli_2005,Ziebell_2005}, we reproduce this case here in order to compare with the case where photoionization is added. We see in Fig.~\ref{fig:z_eps}(a) that the roots corresponding to the whistler branch display decreasing values of $|z_\mathrm{r}|$ for increasing dust density up to a certain point, after that this value starts to increase together with $\epsilon$. The roots corresponding to the ion cyclotron branch also show decreasing value of $|z_\mathrm{r}|$ for increasing $\epsilon$ until a point where they cross each other and then keep their real values practically constant with increasing dust density. Fig.~\ref{fig:z_eps}(b) shows that the damping of both modes increases with dust density up to a certain value of $\epsilon$, after that the damping rates of the whistler branch continues to increase while the ion cyclotron branch is gradually reduced.

When the photoelectric current is added through a stellar temperature of $T_\mathrm{s}=4500$\,K (negatively charged grains) we get the case shown in the mid panels of Fig.~\ref{fig:z_eps}. We see from Fig.~\ref{fig:z_eps}(c) that the whistler and ion cyclotron modes again have decreasing values of $|z_\mathrm{r}|$ for increasing $\epsilon$ up to a certain point, after which they have a different behaviour when compared with Fig.~\ref{fig:z_eps}(a). Now the whistler modes keep its values $|z_\mathrm{r}|$ practically constant while the ion cyclotron modes show increasing values of $|z_\mathrm{r}|$ with increasing dust density. We also notice that the point of $\epsilon$ where the ion cyclotron modes meet is larger when we consider photoionization, and that the whistler modes cross the curves corresponding to the ion cyclotron modes, something that does not happen in Fig.~\ref{fig:z_eps}(a). Fig.~\ref{fig:z_eps}(d) shows that the damping rates of the modes also behave differently when photoionization is considered, in this case it is the ion cyclotron modes that display increasing damping rate for increasing dust density, while the whistler modes gradually reduce its values of $|z_\mathrm{i}|$.

This coupling between different modes and the exchange of behaviours of the damping curves was also observed by \citet{Ziebell_2005} for the case without photoionization, but starting in a greater value of wavenumber, about $q\simeq0.2$, for the same set of parameters. Hence, we conclude that the presence of radiation can reduce the minimum value of wavenumber where the whistler mode crosses the ion cyclotron mode and an exchange on the behaviours of the modes occurs.

The case of positively charged grains ($T_\mathrm{s}=5000\,\text{K}$) can be seen in the bottom panels of Fig.~\ref{fig:z_eps}. We notice that now the curves corresponding to the ion cyclotron mode no longer meet as in previous cases. The damping rates of the ion cyclotron mode continue to show increasing values for increasing $\varepsilon$, as in the case of $T_\mathrm{s}=4500$\,K. The whistler waves also present similar behaviour of the damping rates when compared with $T_\mathrm{s}=4500$\,K, with a larger value of $\varepsilon$ where $|z_\mathrm{i}|$ starts to decrease.

\section{Conclusions}
\label{sec:Conclusions}

We have used a kinetic formulation to study the propagation and damping of waves in a homogeneous magnetized dusty plasma with parameters that can be typically found in stellar winds. Dust particles were assumed to be immobile and have variable charge caused by absorption of plasma particles and by photoionization. We have considered the case of propagation of waves and radiation exactly parallel to the ambient magnetic field. Our analysis has been done focusing on the changes that the addition of the photoionization process brings to the dispersion relation, comparing it with previous studies where only the absorption of particles has been considered as charging mechanism of the dust grains.

The results show that the coupling between the whistler and ion cyclotron modes into the Alfvén mode, which is known to occur in the absence of dust for small values of wavenumber, no longer happens in the presence of negatively charged dust particles, but may occur in a certain region of wavenumber values for neutral or positive grains. We also see in Fig.~\ref{fig:zr_q_small} that these modes show null group velocity in a interval of small wavenumber values when dust is present in the plasma, and that the photoionization process tends to reduce the maximum value of wavenumber where the waves are non-propagating.

The study of the damping rates of the whistler and ion cyclotron modes for small value of dust density shows that these modes present a new damping mechanism caused by inelastic collisions, which the photoionization process tends to attenuate for the parameters considered in Fig.~\ref{fig:zi_q}. For very small values of wavenumber the damping rates of these modes may exchange behaviours when the dust particles' electrical charge changes sign, showing either very small or very high values of damping rates.

The analysis of the effects of different temperatures of the plasma species on the damping rates of the whistler and ion cyclotron modes (Fig.~\ref{fig:zi_q_temp}) shows that the reduction of ion temperature by a factor of five will greatly decrease the ion cyclotron damping rates for small values of wavenumber, being this rate almost constant until it reaches the region where the non-collisional damping takes place. The whistler mode also shows a decrease of its damping rate for reduced ion temperature, where it shows increasing damping values for larger wavenumber.

Finally, in Fig.~\ref{fig:z_eps} we investigate the dependence of the frequencies with the dust density for a fixed wavenumber.  We notice that the presence of photoionization can greatly modify the real and imaginary values of the modes for large dust densities. The whistler mode, for example, shows damping values that increase almost linearly with dust density when we consider only absorption of particles as charging mechanism. But, if we add the photoionization process, this mode can decrease its damping rate for increasing values of dust density.

Our results have shown that the presence of a dust component in the stellar wind from carbon-rich stars substantially alters the dispersion characteristics and wave-particle interactions of the Alfvén waves generated by fluctuations of the ambient magnetic field. Since it is believed that Alfvén waves play a significant role in the heating and acceleration processes that take place in the wind, our results have shown the need for further research on the subject. 

There are several avenues of investigation that can be followed from our work. One of such avenues is the study of the effects of dust on the dispersion and absorption of oblique-propagating Alfvén waves, since the electric field of these waves, particularly of the kinetic Alfvén waves, can operate as an additional acceleration force for the plasma particles. We are currently pursuing this subject and intend to publish our findings in the near future.

However, a more complete investigation about the effects of the dust on the propagation and damping of Alfvén waves occurring in stellar winds, and their subsequent influence on the dynamical evolution of the wind must necessarily be carried out by a theoretical model that takes into account the nonlinear nature of the local wave-particle and wave-wave interactions. The present work is based on the linearized form of the Klimontovich-Maxwell system of equations and the results we obtained can be thus considered as initial conditions for a more comprehensive treatment that takes into account the higher-order nonlinear terms in the Klimontovich-Maxwell system.  Such treatment is a logical and necessary extension of our work that we intend to pursue in the future.  Notwithstanding, from the observed and inferred influence of Alfvén waves on the solar wind, it is expected that the modifications that occur due to the dust population reported here will ultimately affect the nonlinear processes where Alfvén waves are involved in the dust-rich environment of carbon stars, thereby affecting their processes of coronal heating and wind acceleration and energization.

Our formalism can also be applied to other astrophysical environments where a magnetized plasma is contaminated by dust particles. The magnetospheres of Jupiter and Saturn are obvious examples, since the composition of the magnetospheres of these giants is constantly affected by geologically-active satellites and by planetary rings. In such environments, the charge of the dust is observed to vary substantially within the magnetosphere, thereby rapidly altering the characteristics of the Alfvén waves. 

\section*{Acknowledgements}

This study was financed in part by the Coordenação de Aperfeiçoamento de Pessoal de Nível Superior – Brasil (CAPES) – Finance Code 001. LBdT acknowledges support from CNPq (Brazil), grant No. 130965/2019-7. RG acknowledges support from CNPq (Brazil), grant No. 307845/2018-4.

\section*{Data Availability}

The data underlying this article will be shared on reasonable request to the corresponding author.



\bibliographystyle{mnras}
\bibliography{MAIN.bib} 




\appendix

\section{Component \texorpdfstring{$\varepsilon^\mathrm{P}$}{} for parallel propagation}

Considering a population of dust particles with constant radius $a$ and radiation propagating in parallel to the ambient magnetic field we may expresses the component of the dielectric tensor related to the photoelectric current as \citep{galvao_ziebell2012}
\begin{equation}
    \epsilon_{ij}^\mathrm{P}=U_{i}^\mathrm{P}S_{j}^\mathrm{P},
\end{equation}
where
\begin{equation}
    \begin{aligned}
    U_{i}^\mathrm{P}=&-\frac{1}{\omega+\mathrm{i}\nu_{ch}+\mathrm{i}\nu_{1}-\mathrm{i}\nu_{P1}}\frac{e}{am_{e}^{2}}\sum\limits _{l=-\infty}^{\infty}\int \mathrm{d}^{3}p~p_{\parallel}\\
    &\times \frac{p_{\parallel}^{\delta_{i3}}p_{\perp}^{\delta_{i1}+\delta_{i2}}\sigma_\mathrm{p}'(p)F(p)}{\omega-l\Omega_{e}-\frac{k_{\parallel}p_{\parallel}}{m_{e}}+\mathrm{i}\nu_{ed}^{0}(p)} \Pi_{i3}^{le}H(p_{z}) ,
    \end{aligned}
    \label{eq:U^P}
\end{equation}
\begin{equation}
    \begin{aligned}
    S_{j}^\mathrm{P} =&-a \sum\limits_{l=-\infty}^{\infty} \sum\limits_{\beta} \frac{\omega_{p\beta}^{2}m_{\beta}}{n_{\beta0}} \int \mathrm{d}^{3}p\frac{[\nu_{\beta d}^{0}/\omega]\Pi_{3j}^{l\beta}}{\omega-l\Omega_{\beta}-\frac{k_{\parallel}p_{\parallel}}{m_{\beta}}+\mathrm{i}\nu_{\beta d}^{0}} \\
    & \times \left(\frac{p_{\parallel}}{p_{\perp}}\right)^{\delta_{j3}} \left[\mathcal{L}(f_{\beta0})+\mathrm{i}\delta_{j3}\frac{\nu_{\beta d}^{0}(p)}{\omega}\frac{L_{\beta}(f_{\beta0})}{p_{\parallel}}\right]\\
    & +\delta_{j3}\frac{a}{\omega}\sum\limits _{\beta}\frac{\omega_{p\beta}^{2}m_{\beta}}{n_{\beta0}}\int \mathrm{d}^{3}p~\frac{\nu_{\beta d}^{0}(p)}{\omega}\frac{L_{\beta}(f_{\beta0})}{p_{\perp}},
    \end{aligned}
    \label{eq:S^P}
\end{equation}
with
\begin{equation}
    \mathbf{\Pi}^{l\beta}=
    \begin{pmatrix}
        \frac{l^{2}J_{l}^{2}}{b_{\beta}^{2}} & \mathrm{i}\frac{lJ_{l}^{'}J_{l}}{b_{\beta}} & \frac{lJ_{l}^{2}}{b_{\beta}}\\
        -\mathrm{i}\frac{lJ_{l}^{'}J_{l}}{b_{\beta}} & J_{l}^{'2} & -\mathrm{i}J_{l}^{'}J_{l}\\
        \frac{lJ_{l}^{2}}{b_{\beta}} & \mathrm{i}J_{l}^{'}J_{l} & J_{l}^{2}
    \end{pmatrix}
\end{equation}
where $b_{\beta}=k_{\perp}p_{\perp}/m_{\beta}\Omega_{\beta}$ is the Larmor radius and $J_{l}=J_{l}(b_{\beta})$ are the Bessel functions of the first kind. We point out to the work of \citet{galvao_ziebell2012} for the explicit forms of others terms in equations~\eqref{eq:U^P} and \eqref{eq:S^P}, since they are not of interest in the present work.

Using a small Larmor radius expansion for the Bessel functions and their derivatives
\begin{equation}
    J_{l}\simeq\frac{b_{\beta}^{l}}{2^{l}(l!)},\quad J_{l}^{'}\simeq\frac{lb_{\beta}^{(l-1)}}{2^{l}(l!)},
\end{equation}
is easy to show that, for parallel propagation ($k_{\perp}=0$), we have $\Pi_{i3}^{l\beta}=0$ for $i\neq3$ and $\Pi_{3j}^{l\beta}=0$ for $j\neq3$, resulting in
\begin{equation}
    \mathbf{\epsilon}^\mathrm{P}=
    \begin{pmatrix}
        0 & 0 & 0\\
        0 & 0 & 0\\
        0 & 0 & U_{3}^\mathrm{P}S_{3}^\mathrm{P}
    \end{pmatrix},
    \label{eq:tensor_photo}
\end{equation}
which shows that in this case the component $\epsilon^\mathrm{P}$ of the dielectric tensor does not contribute to the dispersion relation for Alfvén waves.


\bsp	
\label{lastpage}
\end{document}